\documentclass{article}

\usepackage{times}
\usepackage{latexsym}
\usepackage{tikz}
\usepackage{pgfplots}
\usepackage{tikz}
\usetikzlibrary{patterns}
\usepgfplotslibrary{groupplots}
\usepackage{tabularx}
\usepackage{amsmath,amsthm,mathtools,tensor,bm,cancel}
\usepackage{amssymb}
\usepackage{caption}
\usepackage{subcaption}
\usepackage[inline]{enumitem}
\usepackage{xspace}
\usepackage{comment}
\usepackage{floatrow}
\usepackage{caption}
\usepackage{subcaption}

\newfloatcommand{capbtabbox}{table}[][\FBwidth]

\usepackage{float}
\restylefloat{table}

\newcolumntype{L}{>{\RaggedRight\arraybackslash}X}

\newtheoremstyle{tightstyle}
  {2pt} 
  {2pt} 
  {} 
  {} 
  {\bfseries} 
  {.} 
  {.5em} 
  {} 

\theoremstyle{tightstyle}

\theoremstyle{tightstyle} 
\theoremstyle{tightstyle}

\newcommand\base{\texttt{Base}\xspace}
\newcommand\stdfine{\texttt{Finetuned}\xspace}
\newcommand\oldreg{\texttt{OutReg}\xspace}
\newcommand\newreg{\texttt{ITVReg}\xspace}
\newcommand\newaug{\texttt{ITVAug}\xspace}

\newcommand\simcse{\texttt{SimCSE}\xspace}
\newcommand\maskreg{\texttt{MaskReg}\xspace}

\renewcommand{\paragraph}[1]{\noindent \textbf{#1}}

\usepackage{array,multirow}

    \usepackage[final]{neurips_2020}

\usepackage[utf8]{inputenc} 
\usepackage[T1]{fontenc}    
\usepackage{hyperref}       
\usepackage{url}            
\usepackage{booktabs}       
\usepackage{amsfonts}       
\usepackage{nicefrac}       
\usepackage{microtype}      

\title{Improving Out-of-Distribution Generalization of Text-Matching Recommendation Systems}

\author{%
  Parikshit Bansal \\
  Microsoft Research \\
  \texttt{parikshitb52@gmail.com} \\
   \And
   Yashoteja Prabhu \\
   Microsoft Research \\
   \texttt{yprabhu@microsoft.com} \\
   \And
   Emre K\i c\i man \\
   Microsoft Research \\
   \texttt{emrek@microsoft.com} \\
   \And
   Amit Sharma \\
   Microsoft Research \\
   \texttt{amshar@microsoft.com} \\
}

\begin{document}

\maketitle

\begin{abstract}
Given a user's input text, \textit{text-matching} recommender systems output relevant items by comparing the input text to available items' description, such as product-to-product recommendation on e-commerce platforms.  
As users' interests and item inventory are expected to change, it is important for a text-matching system to generalize to data shifts, a task known as out-of-distribution (OOD) generalization.
However, we find that the popular approach of fine-tuning a large, base language model on paired item relevance data (e.g., user clicks) can be counter-productive for OOD generalization. 
For a product recommendation task, fine-tuning  obtains \textit{worse} accuracy than the base model when recommending items in a new category or for a future time period. 
To explain this  generalization failure, 
we consider an \textit{intervention-based} importance metric, which shows that  a fine-tuned model captures spurious correlations  and 
fails to learn the causal features that determine the relevance between any two text inputs. 
Moreover, standard methods for causal regularization 
do not apply in this setting, because
unlike in images, there exist no universally spurious features in a text-matching task (the same token may be spurious or causal depending on the text it is being matched to).  
For OOD generalization on text inputs, therefore, we highlight a different goal: avoiding high importance scores for certain features. 
We do so using an {intervention-based} regularizer that constraints the importance score of any token on the model's relevance score to be similar to the base model.  
Results on Amazon product  and 3 question recommendation datasets 
show that 
our proposed regularizer improves generalization for both in-distribution and OOD evaluation, especially in difficult scenarios when the base model is not accurate. 
\end{abstract}

\section{Introduction}
In item-to-item recommendation~\cite{adomavicius2005toward}, the goal is to output a list of relevant items given an input item. Many such recommender systems utilize the text content  of an item, such as recommending relevant questions given a question in an online forum~\cite{wang2018glue}, suggesting relevant products given a product title~\cite{linden2003amazon}, or predicting relevant ads given a search engine query~\cite{broder2008search}. A popular way for training these \textit{text-matching} recommender systems is to fine-tune text embeddings from a \textit{base} language model like BERT using supervised user feedback (such as clicks). For example, one may use a contrastive loss to ensure that given an input query, embedding of another user-labelled relevant item is closer to it than any other item~\cite{dahiya2021siamesexml,xiong2020approximate}; or a standard supervised loss to ensure that embeddings of items within each labelled pair are similar~\cite{reimers2019sentence}.
Pairwise distances on the learnt embeddings are then used to rank candidate items for recommendation, in a bi-encoder architecture~\cite{reimers2019sentence}.

However, the generalizability of the fine-tuned embeddings to item distributions beyond the user-labelled data has received little attention. Distribution shifts are common in recommendation systems, such as when new product categories are added to an e-commerce platform, the list of recommendable items is modified, or the popularity of items changes over time.  As a result, after a recommender model is deployed, it is likely to encounter out-of-distribution data compared to its training set.

For out-of-distribution generalization, we find that fine-tuned models using  relevance labels
can be worse than the pre-trained base model that they start from.  On a product recommendations dataset from Amazon, while fine-tuning always increases in-distribution accuracy, the accuracy of the fine-tuned model on unseen product categories is lower than that of the base model.  We find a similar result on question-to-question recommendation on online forums (\textit{sentence matching} task~\cite{reimers2019sentence,thakur2020augmented}). Training on one forum's data and evaluating on another can result in lower accuracy compared to the base model, especially when the forums are farther apart in content (e.g., general-purpose \textit{Quora} versus technical forums like \textit{AskUbuntu} and \textit{SuperUser}). 



To improve generalizability, we consider a causal graph for the relevance function's computation and highlight the difficulty of not having universally spurious tokens for the text-matching task. Instead, we argue that avoiding disproportionately high importance scores for tokens can be a viable way to regularize for OOD generalization. Specifically, we propose \newreg, a method that constrains the importance of tokens to be similar to that in the base model, which has been trained on a larger and more diverse set and thus less likely to share the same spurious patterns. To do so, the method utilizes \textit{interventions} on the training data (e.g., by masking certain tokens) to create OOD queries. We also provide a  data augmentation version of our method for computational efficiency; and a simpler regularizer (\oldreg) that makes the model's output equal to the base model's output.

We evaluate \newreg on OOD data for product title and question  recommendation tasks  and find that it improves OOD accuracy of naively finetuned models. Compared to a baseline of directly regularizing a model's predictions to the base model, \newreg performs the best under reasonable OOD shifts, where exploiting training data is useful. Interestingly,  
\newreg also helps accuracy on an IID test set: it  increases accuracy over low-frequency, ``tail'' items.
 Our contributions include,   
\begin{itemize}[leftmargin=*,topsep=0pt]{}
    \setlength\itemsep{0.5pt}
    \item When new items (of a different category or from a later time) are introduced, state-of-the-art models for text-matching recommendation are worse than the base model on which they are fine-tuned.
    \item We characterize how new items break correlation between tokens of the query and candidate items; thus methods overfitting to those correlations fail.
    \item We propose a regularizer based on an interventional measure of token importance that improves OOD generalization  on two benchmark tasks.
\end{itemize}

\section{Related Work}
Our work connects recommender systems, sentence matching, and OOD generalization literature.

\paragraph{Text-matching recommendation systems.}
Among item-to-item recommendation systems, text-matching is a popular technique since many items can be characterized through their text,  e.g., product-to-product recommendation on e-commerce websites or question-to-question recommendation on online forums. Given an input query (such as a product's title or description), the goal is to find the most relevant items. State-of-the-art models use dense text retrieval techniques~\cite{xiong2020approximate,karpukhin2020dense}, based on a pre-trained \textit{base model} like BERT~\cite{devlin2018bert} for the initial encoding. Then, either a bi-encoder or cross-encoder architecture~\cite{thakur2020augmented,karpukhin2020dense} is trained for learning the similarity between any two items. The former is computationally efficient while the latter is more accurate but inefficient~\cite{reimers2019sentence}. Since recommender systems often deal with a large number of items, we restrict ourselves to bi-encoders.  Contrastive loss with negative mining~\cite{yu2022selfsupervisedsurvey,dahiya2021siamesexml,hoang2022learning} is a popular way to train a bi-encoder model because  recommender systems usually have one-sided feedback on the relevant pairs of items.


In addition to having a short text description about the item, recent work augments training data with longer descriptions~\cite{dahiya2021siamesexml} or user-item graphs~\cite{mittal2021eclare,saini2021galaxc} to help improve generalization and deal with new users or items.  However, we focus on single sentence inputs since they are most common and generalize to almost all domains. 

A major concern for recommender systems is the spurious correlations present in training data (e.g., exposure bias~\cite{schnabel2020debiasing} or popularity bias~\cite{abdollahpouri2019unfairness}) and how to avoid them for generalization to new kinds of items or users. While these questions have been explored for user-item recommendation models~\cite{zhou2021contrastive,he2022causpref}, text-based item-to-item models have received less attention. 





\paragraph{Sentence matching.}
Recent work on  sentence matching task has focused on improving model architectures for better in-distribution accuracy: efficient bi-encoders instead of cross-encoders~\cite{reimers2019sentence} or using a different architecture than BERT~\cite{gao2021condenser}. Others have focused on buiding a better base model, by self-supervision on  natural language inference datasets~\cite{lee2019latent,gao2021simcse,chang2020pre} or utilizing the stochasticity in model dropout to create training pairs~\cite{gao2021simcse}.    
Closer to our work, \cite{thakur2020augmented} tackled the problem of domain adaptation by assuming access to unlabelled data from a new distribution, and leveraging a cross encoder to weakly label that data. Rather than domain adaptation, our work focuses on the harder domain generalization problem~\cite{zhou2021domain} where no data is available {\em a priori} from the new domains.  


\paragraph{OOD generalization and fine-tuning.} Domain generalization in the vision literature~\cite{zhou2021domain}  aims to identify spurious features in image data and remove them from a model's representation using data augmentation on the spurious feature~\cite{yao2022improving} or through regularization~\cite{cha2022domain}. While recent work has attempted using data augmentations from vision~\cite{xie2020contrastiveaugs}, such augmentations are not always useful in text data~\cite{yugraphaugcritique2021} since it is difficult to obtain universal augmentations. For instance, in the recommendation scenario, certain tokens (e.g., brand of a product) may be both spurious and causal depending on the user intent (e.g., substitute or accessory for a product).

Given the prevalance of pre-trained base models, another direction is to utilize the base models for OOD generalization since they are trained on larger, diverse data ~\cite{hendrycks2020pretrained}. Recent work proposes  fine-tuning-based OOD generalization on image classification~\cite{kumar2022fine,wortsman2021robust,cha2022domain}, but this direction has not been explored for text models or contrastive loss,  especially in the  recommendations context. 





\section{Distribution shifts in recommendation}
\subsection{Review of SOTA recommendation models}
\label {sec:review_sota}
Let $\mathcal{X}$ be the set of input queries and $\mathcal{Z}$ be the set of candidate items. Let the train data be ${(X_i,Z_i,R_i)}_{1\ldots N}$, where, $X_i,Z_i$ correspond to the query and item text respectively and $R_i$ is the relevance or similarity score between $X_{i},Z_{i}$. For each query $X$, the text-matching recommendation problem is to output a ranked list of $k$ items $\mathcal{Z}_k \subset \mathcal{Z}$. 

\paragraph{Encoder.} We define $f_{\theta} : \mathcal{X} \rightarrow \mathbb{R}^N $ as our encoder, parameterised by $\theta$. We use $\theta_0$ as parameters for \base model and hence $f_{\theta_0}(.)$ as the \base encoder. We assume that $f$ outputs unit norm embeddings, i.e. $||f_{\theta}(X)||_2 = 1$ and $||f_{\theta_0}(X)||_2 = 1$. 

\paragraph{Training.} For training the recommendation model, state-of-the-art methods~\cite{dahiya2021siamesexml,xiong2020approximate}
embed both queries and items in a common space using a large, pre-trained base encoder like BERT that provides the initial embeddings.  Thus, we can assume pre-trained base embeddings $f_{\theta_0}(X)$  and $f_{\theta_0}(Z)$ for each query and item  respectively. These embeddings are then fine-tuned using a supervised loss on the paired relevance labels. A popular loss is the contrastive loss~\cite{lee2021ance,dahiya2021siamesexml}
based on sampling negative labels from the query-item pairs that are not labelled as relevant by users. The intuition is to bring closer the embeddings of the relevant pairs while pushing away the embeddings of the non-relevant pairs. 
\begin{equation}\label{eq:contrloss}\small
    \mathcal{L} = \max(0, 1+f_{\theta}(X)^{\top}f_{\theta}(Z^-)
    -f_{\theta}(X)^{\top}f_{\theta}(Z^+))
\end{equation}
where $X$, $Z^+$ and $Z^-$ denote a query, relevant item, and a non-relevant item respectively; and the relevance score of any query-item pair is given by the cosine similarity of their embeddings (equal to the dot product for unit norm). When ratings or relevance scores are available, one may also use the MSE loss~\cite{reimers2019sentence}.

\paragraph{Inference.} 
For a new input query, the model outputs a vector $\hat{R} \in \mathbb{R}^M$ where $M$ is  number of items, and $r_j = f_{\theta}(X)^{\top} f_{\theta}(Z_j)$ is its \textit{relevance score} with $j$th item. The $k$ items with the highest relevance scores,  $\mathcal{Z}_k$ are output as recommendations. 


\paragraph{Evaluation.} The system is typically  evaluated on its accuracy in predicting the ``ground-truth'' relevance labels (e.g., the metric may be precision over the top-k ranked items). The evaluation data is a held-out random sample of the queries from the same log data as training. 
This evaluation scheme, however, assumes that the queries and items are \textit{independently and identically distributed} (IID), at training and  deployment time. In practice, this is rarely the case. The distribution of queries $P(X)$ from users  changes over time, in predictable (e.g., seasonal) and unpredictable ways. Similarly, the distribution of candidate items $P(Z)$ changes based on item availability (e.g., as new item categories are added to a system). Therefore, it is important to evaluate a recommendation model not just on IID, but also on out-of-distribution (OOD) samples. Ideally, the model should obtain high performance on both  IID and OOD samples. 

\begin{figure}
\begin{floatrow}
\ffigbox{
     \centering
     \begin{subfigure}[b]{0.40\columnwidth}
         \centering
         \includegraphics[width=\textwidth]{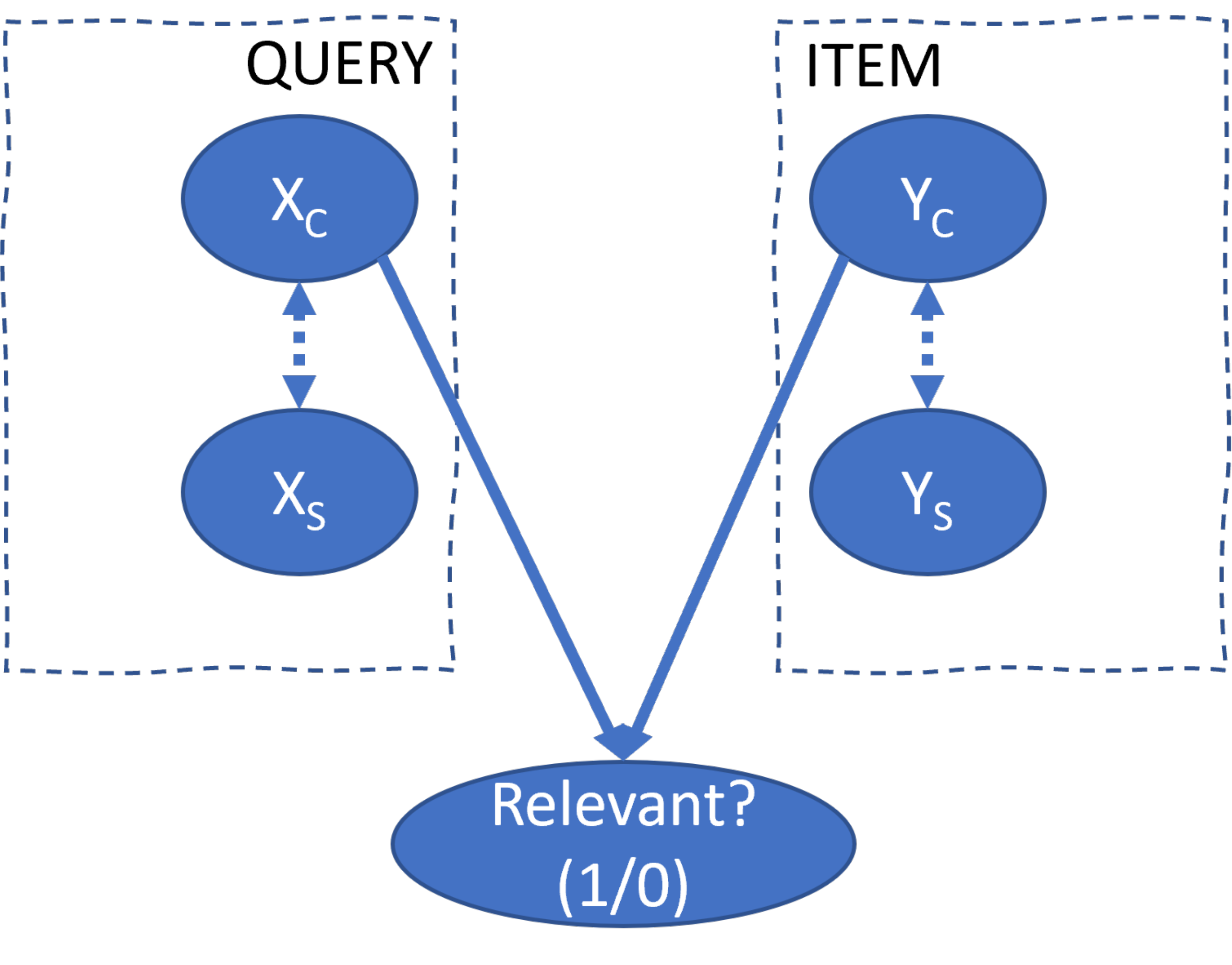}
         \caption{General graph.}
         \label{fig:y equals x}
     \end{subfigure}
     \begin{subfigure}[b]{0.52\columnwidth}
         \centering
        \includegraphics[width=\textwidth]{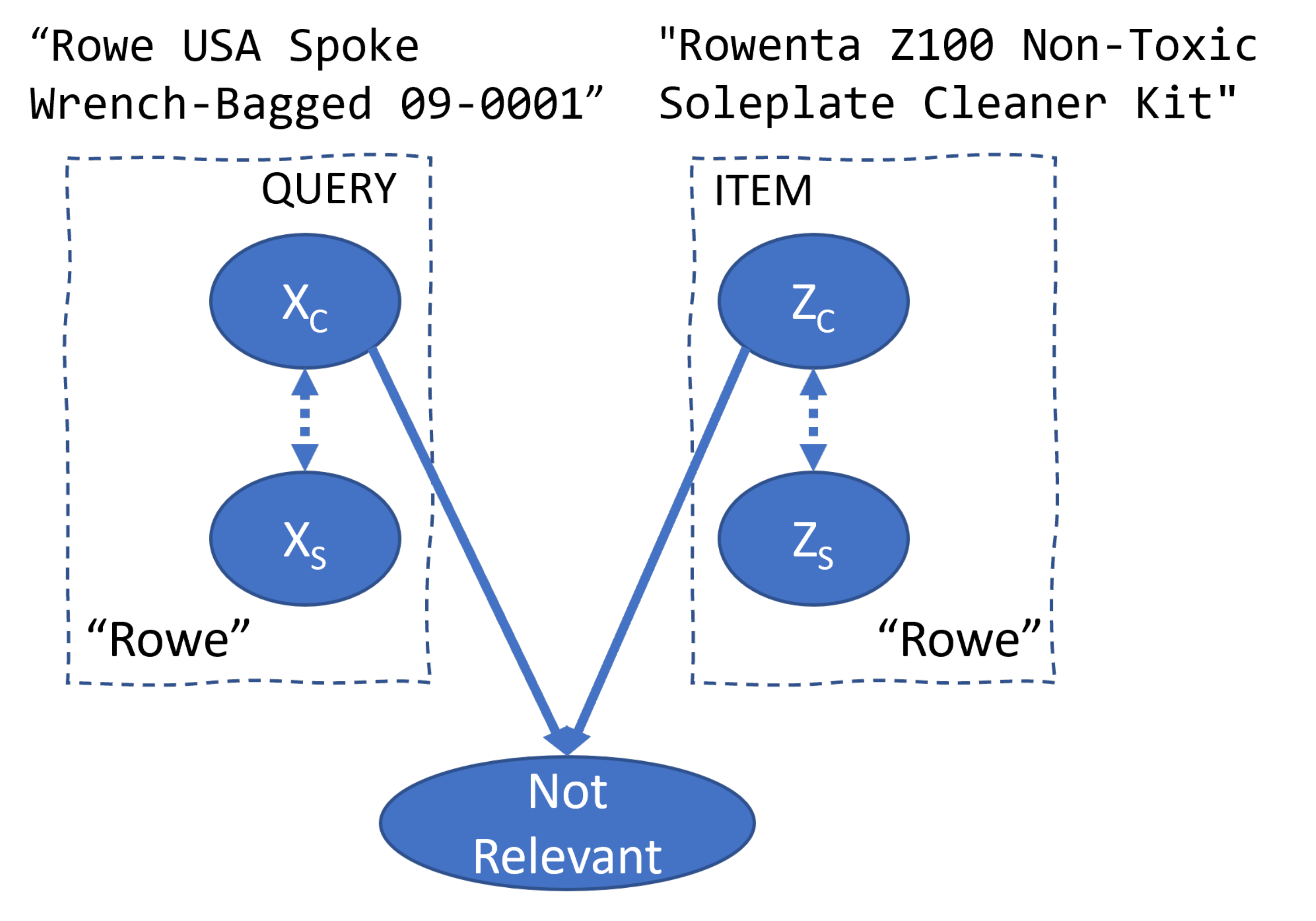}
        \caption{Example graph.}
     \end{subfigure}}
     {\caption{Causal graph for the data-generating process for query, item and relevance score, $(X,Z,R)$. Spurious tokens $X_s$ or $Z_s$ do not cause query-item  relevance but are predictive of the relevance score.} \label{fig:graph_ql}}

\capbtabbox{%
    \centering
    \begin{tabular}{c|cc}
    
        Method & IID & OOD  \\
        \hline
        Base Model & 20.01 & \textbf{30.61} \\
        Fine-tuned Model & \textbf{$38.74 \pm 0.08$} &  $28.31 \pm 0.17$.
    \end{tabular}
}
{\caption{Precision@1 for recommending product titles on Amazon. OOD denotes queries from five categories that were not included in the train data. Even after training on a similar dataset of Amazon titles, the precision of the  finetuned model is lower than the base model.}
        \label{tab:ft_vs_base}}

\end{floatrow}
\end{figure}


    

\subsection{Types of distribution shift}
We study three shifts: change in queries $P(X)$, items $P(Z)$, and both  queries and items $P(X,Z)$. 

\paragraph{Change in P(X).} The distribution of queries changes but the set of candidate recommendations may remain the same. 
For example, the popularity of certain queries can shoot up due to external events, or the  system may be expanded to new queries (e.g., products on another website). 

\paragraph{Change in P(Z).} The distribution of candidate items changes while the queries remain the same. For example, an e-commerce platform may alter the eligibility rules  for an item to be recommended, or  recommend items from a partner website.

\paragraph{Change in both P(X) and P(Z).} A final scenario is when the distribution of both queries and items changes. For example, introducing a new item category in an online store 
leads to  new queries that can be accessed by user and also be recommended. 

While the criteria for relevance (i.e., user intent) can change over time, for simplicity, we assume that $P(R| Z, X)$ remains constant where $R$ is the true relevance score.  
As we show below, even under this favorable  assumption, text-matching models can learn patterns that do not generalize. 



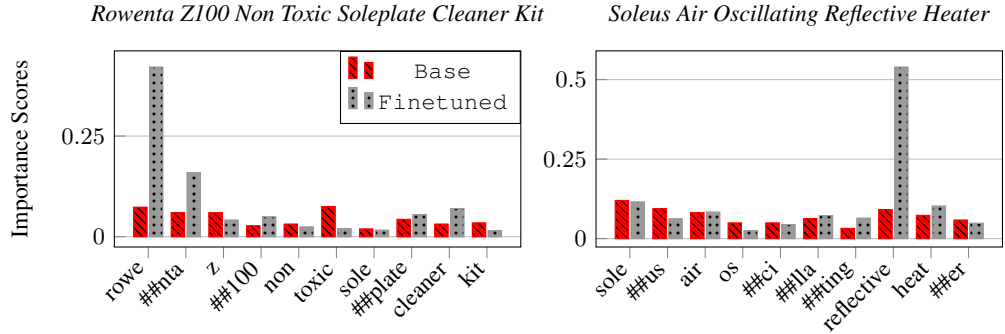
\begin{figure*}
\centering
\begin{tikzpicture}
\begin{groupplot}[group style={group size= 2 by 1},width=0.5\textwidth,height=0.30\textwidth]
\nextgroupplot[ybar=1.0pt, 
tick label style={font=\footnotesize},
symbolic x coords={rowe,\#\#nta,z,\#\#100,non,toxic,sole,\#\#plate,cleaner,kit},
xtick={rowe,\#\#nta,z,\#\#100,non,toxic,sole,\#\#plate,cleaner,kit},
ytick={-0.5,-0.25,0,0.25,0.5},
xtick pos=left,
ytick pos=left,
x tick label style={rotate=45,anchor=east},
scaled y ticks = false,
legend pos = north west, 
ymajorgrids=true,
ylabel = {\footnotesize Importance Scores},
legend style={at={(1,1)},anchor=north east,legend columns=1,font=\footnotesize},
title={\small \textit{Rowenta Z100 Non Toxic Soleplate Cleaner Kit}},
bar width=5pt]
\addplot+[red, fill=red, postaction={pattern=north west lines}] coordinates {(rowe, 0.073) (\#\#nta, 0.060) (z, 0.060) (\#\#100, 0.027) (non, 0.031) (toxic, 0.075) (sole, 0.019) (\#\#plate, 0.043) (cleaner, 0.031) (kit, 0.034)  };
\addplot+[black!40!white, fill=black!40!white, postaction={pattern=dots}] coordinates {(rowe, 0.421) (\#\#nta, 0.159) (z, 0.041) (\#\#100, 0.049) (non, 0.024) (toxic, 0.020) (sole, 0.016) (\#\#plate, 0.055) (cleaner, 0.070) (kit, 0.015)};

\legend{\base, \stdfine};
\coordinate (c1) at (rel axis cs:0,1);

\nextgroupplot[ybar=1.0pt, 
tick label style={font=\footnotesize},
symbolic x coords={sole,\#\#us,air,os,\#\#ci,\#\#lla,\#\#ting,reflective,heat,\#\#er},
xtick={sole,\#\#us,air,os,\#\#ci,\#\#lla,\#\#ting,reflective,heat,\#\#er},
ytick={-0.5,-0.25,0,0.25,0.5},
xtick pos=left,
ytick pos=left,
x tick label style={rotate=45,anchor=east},
scaled y ticks = false,
legend pos = north west, 
ymajorgrids=true,
legend style={at={($(0,0)+(1cm,1cm)$)},legend columns=1,fill=none,draw=black,anchor=center,align=left,legend cell align=left,font=\small},
title={\small \textit{Soleus Air Oscillating Reflective Heater}},
bar width=5pt]
\addplot+[red, fill=red, postaction={pattern=north west lines}] coordinates {(sole, 0.120) (\#\#us, 0.094) (air, 0.082) (os, 0.049) (\#\#ci, 0.049) (\#\#lla, 0.063) (\#\#ting, 0.032) (reflective, 0.091) (heat, 0.073) (\#\#er, 0.058)  };
\addplot+[black!40!white, fill=black!40!white, postaction={pattern=dots}] coordinates {(sole, 0.115) (\#\#us, 0.062) (air, 0.083) (os, 0.025) (\#\#ci, 0.044) (\#\#lla, 0.072) (\#\#ting, 0.064) (reflective, 0.539) (heat, 0.102) (\#\#er, 0.048)  };

\coordinate (c2) at (rel axis cs:1,1);

\end{groupplot}

\coordinate (c3) at ($(c2)$);

\end{tikzpicture}
\vspace*{-10pt}

\caption{\label{exp:both_examples}Token-wise importance scores (Eqn \ref{eqn:importance}) for two example queries 
using the \base and \stdfine models. \base gives approximately equal importance scores to all tokens for both sentences whereas \stdfine model gives dispropotionately high weights to the tokens \textit{"rowe"} and \textit{"reflective"} in first and second sentence respectively.}
\end{figure*}

\subsection{Example: OOD data on Amazon products}
\label{sec:amz-eg}
Let us use the \textit{Amazon Titles} dataset (see Sec.~\ref{sec:amz-data}) to illustrate the problem with OOD generalization for fine-tuned models. Both  the input query and candidate items are titles of products on  Amazon.com. 
We simulate a scenario where $P(X)$ is changed by  removing queries from five  categories from the train data and evaluating the model on those removed queries using  precision@1 metric.

Table~\ref{tab:ft_vs_base} compares the precision@1 of the base model (MSMarco
\cite{sanh2019distilbert}
) and fine-tuned model that is initialized with the base model. As expected,  the fine-tuned model improves significantly over the base model on IID test data, almost doubling the precision. However, on the queries from the unseen item categories, the finetuned model performs  \textit{worse} than the base model.  


\subsection{Explaining the OOD generalization failure}
\label{sec:qual}
\paragraph{Explanation through causal graph.} 
To understand the failure, consider the schematic causal diagram for the data-generating process of the training set. Figure~\ref{fig:graph_ql} shows that each query can be broken down into its semantic (causal) component $X_c$ and its spurious component $X_s$. Same for candidate items, denoted by $Z_c$ and $Z_s$. The semantic components $X_c$ and $Z_c$ together cause the relevance label. An ideal encoder should only learn the semantic components.  However, since the non-semantic components are correlated with the semantic component and can be easier to learn, a fine-tuned encoder may learn the non-semantic components too. 

Specifically, when new queries are introduced, the correlation between $X_c$ and $X_s$ can be broken, even as $P(R|X,Z)$ remains the same. Hence, $X_s$ is no longer a good predictor of relevance and any model relying on $X_s$ will fail to generalize. For example, for certain queries in the Amazon data, $X_s$ may correspond to the brand of a query product and $X_c$ the rest of its title 
. In the training data, the brand (e.g., ``samsung'') may be correlated with a single product category (e.g., ``smartphone'') and thus an encoder may learn a non-zero weight for the brand to determine the relevance score. However, the correlation may be broken in the evaluation data where the same brand is associated with another product category (e.g., ``refrigerator''). 
We can reason analogously for a change in  $P(Z)$.
 


\paragraph{Explanation through interventional examples.}
OOD generalization failure is more intuitively understood through intervention-based analysis~\cite{ribeiro2020beyond}, where we change parts of a query and inspect the model's prediction.  
To understand which tokens a model learns as important, we define an \textit{importance score} $s$ of each token. To compute $s$ for a token,  we mask that token and compute the dot product similarity of the masked input's embedding with the original input's embedding. \cite{bastings2020elephant}
\begin{equation} \label{eqn:importance}
    s_j = 1-f_{\theta}(X)^\intercal f_{\theta}(X_{-j}')
\end{equation}

where $X_{-j}'$ is $X$ with its $j$th token masked. 

Consider a candidate item in the Amazon dataset, "\textit{Rowenta Z100 Non Toxic Soleplate Cleaner Kit}". Figure~\ref{exp:both_examples} (left) shows the importance scores. The base encoder gives almost equal importance to each token, whereas the finetuned encoder is biased towards the token \textit{rowe}, possibly because Rowenta products tend to be relevant to other Rowenta products. This token obtains such a disproportionely high importance score that the item's representation (average of all tokens' representation) is defined by it. Here \textit{rowe} $\in Z_s$ is a spurious token since relevance may depend on the brand \textit{rowenta}, but not on \textit{rowe} alone. When we consider an OOD query having an actual brand called ``Rowe'', "\textit{Rowe USA Spoke Wrench - Bagged 09-0001}", 
the model matches it to the \textit{Rowenta} item and other products by Rowenta (see Suppl. Table~\ref{tab:qualitative_rowe}), exposing the spurious correlation learnt by the encoder. 



As another example, consider the query,  ``\textit{Soleus Air MS-09 Oscillating Reflective Heater}''. The importance scores are in Figure~\ref{exp:both_examples} (right). Again, we see that the base encoder gives almost equal importance to all tokens, while the finetuned encoder is biased towards the token \textit{reflective}. 
This happens since \textit{reflective} is a strong matching signal in the train set: products with this token are often marked as relevant to products that have it too. 
As a result, the top predicted item by the finetuned model is \textit{3M Scotchlite Reflective Tape, Silver, 1-Inch by 36-Inch} (not a relevant item, see Suppl. Table~\ref{tab:qualitative_reflective}). 

\section{Regularization for OOD generalization}
The above analysis indicates that fine-tuning encourages a model to learn high importance for certain tokens while forgetting the rest, which is undesirable if the tokens are spuriously associated with the relevance label. However, it is non-trivial to identify the spurious versus semantic tokens. 
Below we discuss this fundamental limitation and propose two regularizers using the  base model. 


\subsection{Building a Risk Invariance Predictor}
As we mentioned in Section 3, in text-matching systems, we expect the distribution of queries P(X), distribution of labels P(Z), or both to change from train to test data. At the same time, we assumed that $P(R|X, Z)$, i.e.,  the relevance  between a query and label remains invariant across distributions.  That is, the relevance function $g(.|X,Z)$ (for bi-encoders it can be written as $f(X)^\top f(Z)$, see Sec~\ref{sec:review_sota}) remains invariant. Intuitively, the goal is that the relevance function has similar accuracy on train as well as test data.  We can now formally express this intuition of a  OOD genrealizable relevance function, using the definition of risk invariance from \cite{makar2022causally}.  Let $\mathcal{P}$ be a set of distributions over $R,X, Z$ such that $P(R|X,Z)$ remains invariant but the marginal distributions $P(X,Z)$ can vary across the distributions. Then, an optimal predictor can be characterized as follows.

\paragraph{Optimal Risk Invariant Predictor} : Define the risk of predictor $g$ on distribution $P_t \in \mathcal{P}$ as $R_{P_t} (g) = \mathbb{E}_{x,z,r \sim P_t} l(g(x,z), r)$ where $l$ is the loss function. Then, the set of risk-invariant predictors obtain the same risk across all distributions $P_t \in \mathcal{P}$, and set of the optimal risk-invariant predictors is defined as the risk-invariant predictors that obtain minimum risk on all distributions.

From the causal graph of Figure 1,  we assumed that each query can be broken down into two mutually exclusive subsets of tokens, causal $X_c$ and spurious $X_s$ where $X_c \bigcup X_s = X$, such that only $X_c$ affects the relevance with a label. Similarly, label $Z$ can be broken down into two mutually exclusive subsets of tokens, causal $Z_c$ and spurious $Z_s$, such that only $Z_c$ affects the relevance with  a query. Thus, an intuitive way to achieve risk invariance is to build a predictor $g$ for relevance that only depends on $X_c$ and $Z_c$ for any query and label, $P(R|X_c, Z_c)$.

From OOD generalization literature on images~\cite{zhou2021domain,wang2022generalizing}, the goal then is use one of these methods to learn $X_c$/$Z_c$ from observed data: regularization~\cite{mahajan2021domain, lee2022diversify}, weighting~\cite{sagawa2019distributionally,yao2022improving} or data augmentation~\cite{zhou2021domain}.
However, such a solution relies on assumptions that the spurious features (e.g., image background) \textbf{1)} are universal  for all inputs; \textbf{2)} can change independently of the semantic features (e.g., object in image). As we describe below, spurious features in text-matching data do not exhibit such a property. 

\subsection{Problem: Spurious features depends on context}
In text-based recommendations, it is difficult to find (a set of) tokens that are universally spurious, since \textit{spuriousness} depends on  context. For example, in our example with  ``samsung'' products (``smartphone'' and ``refrigerator''), the brand was a spurious feature for learning the relevance score. However, when looking for a smartphone accessory, the brand is no longer spurious (in fact, it is the semantic feature). Thus, the subset $X_c$ for a query may change based on the label, and similarly, the subset $Z_c$ for a label may change based on different queries.  
The second assumption  that tokens can be changed independently is also less plausible. 
Possibly spurious tokens like ``samsung'' can be associated with other parts of the query (e.g, product code or name); it is hard to remove them.

Thus, while there are distribution shifts in text-based queries and items, it is difficult to find universal spurious features that can be modified.  As a result, prior work~\cite{gao2021simcse} has found that methods based on universal data augmentations (e.g., masking certain parts of query, adding a random word) do not work (as we also show in our evaluation). 

Therefore, fully removing the influence of any token can be counter-productive. Rather than  removing the influence of certain tokens, our goal is to ensure that no token is under-weighted such that it ceases to be important (ignored) for relevance prediction. Since we saw that finetuning on text-matching models tend to assign very high weights to certain tokens, we can reframe our goal: to avoid  
learning  large importance scores for some tokens that may lead to ignoring other tokens, and thus not generalize to new data. 

How do we decide a threshold for ``large'' when we regularize importance scores? This is important since we cannot expect each token in a query or label to have the same importance, yet we would like to ensure that tokens do not get disproportionately large importance scores. Rather than choosing arbitrary thresholds on importance scores (e.g., twice than other tokens, K times the mean score, etc.), a good proxy can be the base model. We assume that the base model has been trained on larger, diverse data and is thus unlikely to encode the same spurious patterns. 

We present two methods in this direction below. The first is a brute-force attempt, that forces not just the importance scores but the entire prediction of a model to be the same as base model. The second one is a more targeted regularization that only focuses on the importance scores.

\subsection{Output-based regularization}
Since both the fine-tuned model and the base model have the same architecture, a straightforward way to avoid very high importance scores  is to make the neural network output be the same as the base model. 
In a bi-encoder architecture for text-matching, we can implement this by enforcing that the finetuned and base encoders output exactly the same representation for an input. Given the finetuning $f_{\theta}(.)$ and base $f_{\theta_0}(.)$ encoders and an input query $X$,  
the output regularizer (\oldreg) can be written as, 
$    [f_{\theta}(X) - f_{\theta_0}(X)]^2  $.


This simple regularizer is expected to do well for extreme distribution shifts where learning from IID data may not be that useful, since it constrains the encoder's representation to be closer to the base encoder. 
However, the same property is the biggest weakness of \oldreg. 
As a regularization, it is too strong and discourages learning from train data: with a high enough penalty term, the final encoder may be exactly equal to the base encoder. 

\subsection{Intervention-based regularization}
Therefore, we now describe a weaker notion of base model regularization aimed at less extreme OOD shifts. 
In Section~\ref{sec:amz-eg}, we defined token importance as the similarity between an original query and its modified query with the  token masked. More generally, 
given a text input, we define an \textit{intervention} as an independent change to a subset of the tokens without affecting the rest of the input. The independent change may be any text transformation, such as  masking, deleting, replacing the token subset, etc. By definition, an intervention deviates from the original distribution $P(X)$ that generated the data, thus  creating an out-of-distribution sample. 

For each input $X$, we construct an interventional input $X'$ using a transformation on a random subset $X_{sub}$ of the tokens. The key idea behind our regularizer is that for any subset of tokens $X_{sub}$, its importance score  using the finetuned model should be the same as the importance score using the base model. 
This is a relaxation of the ideal goal of ensuring that the  accuracy of the finetuned model remains the same for $X$ and $X'$. Since we do not have the ground-truth relevance labels for $X'$ (and do not trust the base model to be a good enough proxy for the ground-truth), we  
  regularize on the importance scores.  
Using  Eqn~\ref{eqn:importance} for importance score, the regularizer can be written as, 
\begin{equation} \label{eq:itvreg} \small
\begin{split}
& [(1-f_{\theta}(X)^\intercal f_{\theta}(X')) -     (1-f_{\theta_0}(X)^\intercal f_{\theta_0}(X'))]^2 \\
      &= (f_{\theta}(X)^\intercal f_{\theta}(X')-f_{\theta_0}(X)^\intercal f_{\theta_0}(X'))^2
\end{split}
\end{equation}

We call this the \textit{Interventional} regularizer (\newreg). The full training loss 
(where $L_{ERM}$ can be the contrastive loss from Eqn~\ref{eq:contrloss}) is,
\vspace{-0.2em}
\begin{equation*}\footnotesize
    L_{ERM} + 
     \lambda{\mathbb{E}_X}[(f_{\theta}(X)^\intercal f_{\theta}(X')-f_{\theta_0}(X)^\intercal f_{\theta_0}(X'))^2]
\end{equation*}

\subsection{Equivalent data augmentation procedure}
For computational efficiency, we also provide an augmentation version called \newaug.
From Eqn~\ref{eq:itvreg}, we can write \newreg as ${\mathbb{E}_X}[(f_{\theta}(X)^\intercal f_{\theta}(X')-p)^2]$, 
where $p$ is $f_{\theta_0}(X)^\intercal f_{\theta_0}(X')$ and is a constant throughout the training. Also, $f_{\theta}(X)^\intercal f_{\theta}(X')$ can be interpreted as the predicted relevance between a query $X$ and item $X'$. This is equivalent to adding a data point to training,  $(X_i,X_i',R'_i=f_{\theta_0}(X_i)^\intercal f_{\theta_0}(X_i'))$ for each $i$ and using L2 loss over these augmented points. With $D'$ as the augmented set over $\lambda$ fraction of inputs, we write,  
\begin{equation*} \small
    L_{\newaug} = L_{ERM} + \lambda{\mathbb{E}_{D'}}[(f_{\theta}(X_i)^\intercal f_{\theta}(X'_i)-R'_i)^2]
\end{equation*}

\begin{table*}
\resizebox{\textwidth}{!}{%
\small
\centering
\begin{tabular}{lllll}
\hline
\multirow{2}{*}{Fine-tuning Method} & \multicolumn{2}{c}{Temporal Shift} & \multicolumn{2}{c}{Categorical Shift} \\
& \textbf{Amazon131K (IID)} & \textbf{Amazon1.3M (OOD)} & \textbf{AmazonCatRemoved (IID)} & \textbf{AmazonCatOOD (OOD)} \\
\hline

\base & 22.50 & 25.71 & 20.01 & 30.61  \\\cline{1-5}  
\stdfine & \textbf{39.71 $\pm$ 0.14} & 26.02 $\pm$ 0.08 & \textbf{38.74 $\pm$ 0.08} & 28.31 $\pm$ 0.17 \\\cline{1-5} 
\maskreg & 39.56 $\pm$ 0.01 & 26.65 $\pm$ 0.02 & 37.92 $\pm$ 0.11 & 29.09 $\pm$ 0.06 \\\cline{1-5}
\simcse & 39.47 $\pm$ 0.11 & 26.05 $\pm$ 0.02 & 38.05 $\pm$ 0.56 & 28.52 $\pm$ 0.10 \\\cline{1-5}    
\oldreg & 38.03 $\pm$ 0.53 & \textbf{27.60 $\pm$ 0.03} & 37.66 $\pm$ 0.11 & \textbf{31.21 $\pm$ 0.03} \\\cline{1-5}
\newreg & \textbf{39.72 $\pm$ 0.10} & 27.08 $\pm$ 0.01  & \textbf{38.77 $\pm$ 0.02} & 29.53 $\pm$ 0.04\\\cline{1-5}

\hline
\end{tabular}
}
\caption{P@1 for Temporal and Categorical Shifts on \textit{Amazon Titles}. }
\label{tab:reco}
\vspace*{-5pt}
\end{table*}

\vspace{-0.9em}
\section{Evaluation}
We study OOD generalization of text-matching models on product-to-product and question-to-question recommendation. 
Initial experiments on \newreg indicated that {masking} as the intervention worked the  best, so we use masking for all results. All experiments are run for 3 random seeds and the mean and standard deviation are reported.  For hyperparameter tuning, see Supp.~\ref{sec:hyperparameter_tuning}.

\paragraph{Baselines.} 
We compare \oldreg, \newreg, and \newaug to  the \base model, standard \stdfine model, and  two baselines from past work.
\textbf{1)} \simcse: 
We adapt SimCSE~\cite{gao2021simcse}, an augmentation method for pretraining sentence matching models, for our fine-tuning task.
 During training, we use dropout (seeds $s$, $s'$) to pass an input sentence twice through the model and obtain two embeddings,  considered as a ``relevant'' pair. The regulariser is 
$
(f(X,s)^\intercal f(X,s')  - 1)^2
$.
\textbf{2)} \maskreg: \cite{wu2020clear} propose deleting a span of words in the input as data augmentation. To adapt it into our setup and  compare to \newreg, we  mask a portion of tokens instead. The regulariser is 
$
( f(X)^\intercal f(X')  - 1)^2
$, 
where $X'$ is masked input.

\subsection{Amazon Titles recommendation}
\label{sec:amz-data}


\paragraph{Dataset. } 
We use two product-to-product recommendation datasets, namely \textit{AmazonTitles131K} and \textit{AmazonTitles1.3M}~\cite{Bhatia16}. These datasets contain titles of products that were recommended on a focal product's page. Only the product title can be used for recommendation. There is a time shift though: 131K is collected from the year 2013 and 1.3M is collected from 2014. 
As both new queries and labels are added while moving from 131K to 1.3M, this simulates the $P(X,Y)$ distribution shift setup.
For each dataset, the same candidate items (131K and 1.3M items respectively) are shared across train and validation datasets. Each query in the train set has a labelled set of positive items with relevance $1$. 
The validation set contains unseen queries and we need to recommend the most relevant items for them.



While going from 131K to 1.3M dataset represents a temporal shift where both queries and items' distributions change (and possibly other factors), we simulate a more controlled setup using category information.
We exclude queries from the 5 most 
popular second-level categories 
from the train set, and use them to construct the OOD evaluation set (\textit{AmazonCatOOD}). The in-distribution validation set also has the categories excluded (\textit{AmazonCatRemoved}).  Thus, the item distribution $P(Z)$ remains the same while the query distribution $P(X)$ changes across train and evaluation.
Since $X,Z$ have been treated symmetrically in our setup, this setups results can also be extrapolated to a case where the label distribution $P(Z)$ changes while the query distribution $P(X)$ remains the same.


\begin{figure}[tb]
\centering
\begin{floatrow}
\ffigbox{
\centering
\begin{tikzpicture}
\begin{groupplot}[group style={group size= 1 by 1},width=0.45\textwidth,height=0.35\textwidth]
\nextgroupplot[ybar=1.0pt, 
tick label style={font=\small},
symbolic x coords={0-20,20-40,40-60,60-80,80-100},
xtick={0-20,20-40,40-60,60-80,80-100},
xtick pos=left,
ytick pos=left,
scaled y ticks = false,
legend pos = north west, 
ymin=-6,
ymajorgrids=true,
ylabel = {\footnotesize Percent Gain (\%)},
xlabel = {\footnotesize Item Frequency Quantiles},
legend style={at={(0,0)},anchor=south west,legend columns=2,font=\scriptsize},
bar width=5pt]

\addplot+[brown!40!black,fill=brown!50!white, postaction={pattern=north east lines}] coordinates {(0-20, 3.36909354) (20-40, 0.88327302 )  (40-60,0.24783147) (60-80,-2.27009579)  (80-100, -6.22029732)};
\addplot+[cyan!40!black, fill=cyan!30!white, postaction={pattern=horizontal lines}] coordinates {(0-20,  1.76591128) (20-40, -0.03680304) (40-60, -0.13768415) (60-80, -1.9192309) (80-100, -4.40360529) };
\addplot+[red, fill=red!30!white, postaction={pattern=north west lines}] coordinates {(0-20, 2.94720347) (20-40, 0.79126541) (40-60,  0.52319978)  (60-80,  -1.55082278)  (80-100, -8.50755004)  };
\addplot+[blue, pattern color=blue!40!white, postaction={pattern=crosshatch}] coordinates { (0-20, 3.1219865)  (20-40, 0.90167454)  (40-60,1.21162054) (60-80, -0.09473352) (80-100, -1.67622615) };

\legend{\maskreg,\simcse,\oldreg,\newreg};
\coordinate (c1) at (rel axis cs:0,1);

\end{groupplot}

\end{tikzpicture}
}
{\captionof{figure}{\label{exp:reco_bins}Percent gains in P@1 over \stdfine for different item frequency quantiles. 
\newreg gains on  tail (left) while not losing precision on head (right).}}
\ffigbox{
\centering
\begin{tikzpicture}
\begin{groupplot}[group style={group size= 1 by 1},width=0.45\textwidth,height=0.35\textwidth]
\nextgroupplot[
tick label style={font=\small},
xtick={0,3,6,9,12,15,18,21,24,27},
ytick={-0.5,0},
xtick pos=left,
ytick pos=left,
scaled y ticks = false,
legend pos = north west, 
ymajorgrids=true,
ylabel = {\footnotesize P@1 Difference},
xlabel = {\footnotesize Percentage of 1.3M Items (OOD) Added},
legend style={at={(1.0,0.0)},anchor=south east,legend columns=2,font=\scriptsize },
]
\addplot[brown!40!black,mark=square*] coordinates {(0, -0.139) (3, -0.144) (6, -0.076) (9, -0.025) (12, -0.033) (15, 0.020) (18, 0.050) (21, 0.079) (24, 0.073) (27, 0.100)};
\addplot[cyan!40!black,mark=triangle*] coordinates {(0, -0.342) (3, -0.283) (6, -0.233) (9, -0.173) (12, -0.164) (15, -0.092) (18, -0.103) (21, -0.100) (24, -0.050) (27, -0.062)};
\addplot[red,mark=star] coordinates {(0, -0.699) (3, -0.534) (6, -0.387) (9, -0.217) (12, -0.096) (15, 0.054) (18, 0.111) (21, 0.228) (24, 0.250) (27, 0.348)  };
\addplot[blue,mark=diamond*] coordinates { (0, -0.019) (3, 0.017) (6, 0.055) (9, 0.098) (12, 0.128) (15, 0.154) (18, 0.198) (21, 0.266) (24, 0.254) (27, 0.264) };

\legend{\maskreg,\simcse,\oldreg,\newreg};
\coordinate (c1) at (rel axis cs:0,1);

\end{groupplot}

\end{tikzpicture}
}
{\captionof{figure}{\label{exp:reco_inter} To simulate temporal evolution,  we start with \textit{Amazon131K} and add 39K new items 
from \textit{Amazon1.3M} dataset at each tick.  \newreg is the only method that is consistently better than \stdfine on P@1.}
}
\end{floatrow}
\end{figure}

\paragraph{Fine-tuning.} We follow the modeling procedure used in a recent state-of-the-art work \cite{dahiya2021siamesexml}. We initialize with MSMARCO-DistilBERT-v4 
~\cite{sanh2019distilbert} 
and use contrastive loss with hard negative mining. Details are in Supp~\ref{subsec:hyperparameter_amazon}.

\paragraph{Evaluation metric.} 
We use the $Precision@1$ metric for evaluation. Given predicted similarities $\hat{R} \in \mathbb{R}^M$ and ground truth similarities $R$ (where $r_i$ is 1 if its a relevant item, otherwise 0), $P@k$ is defined as $
P@k = \frac{1}{k}{\sum_{l \in rank_k(\hat{R})}}R_l
$.

\paragraph{Results.}  Table~\ref{tab:reco} shows the IID and OOD P@1 metrics for different 
recommender models. We first analyze the temporal distribution shift: train on \textit{Amazon131K} and test on \textit{Amazon1.3M}. On IID evaluation, \newreg and \stdfine both obtain the highest P@1 while \oldreg obtains the lowest P@1. On OOD evaluation, however, \oldreg obtains the highest P@1 followed by \newreg.

To understand the tradeoff between \newreg and \oldreg, we study how the accuracy of these models would vary as different amounts of OOD data (from 1.3M dataset) is mixed with IID data (131K validation) for evaluation. This simulates the real-world scenario where new items are progressively  added.  Specifically, we progressively add 40K new items and their relevant queries from 1.3M to create multiple OOD datasets. 
As Figure~\ref{exp:reco_inter} shows, \newreg is the only method that performs better than \stdfine (line $y=0$) on P@1 on all OOD datasets. 
As the distribution shift becomes more extreme, (around 24\% of 1.3M OOD data) \oldreg surpasses \newreg 
indicating that \oldreg is more suitable for  large distribution shifts.

Next, we consider categorical distribution shift from Table~\ref{tab:reco}. 
We obtain similar results as the temporal shift:  
%
on the OOD evaluation,  \oldreg followed by \newreg are the best performing models. 
For P@1 on each category, see Suppl. Table~\ref{exp:cat_numbers}.




While we looked at adding new queries and items, in practice, OOD shifts are more commonly observed as  reweighing of existing items. 
Hence, in addition to the average IID accuracy, it is important to analyze accuracy of a model at different levels of item popularity. 
We bin the items according to their frequency in five quantiles i.e. 0-20\%, \ldots, 80-100\%, where lower quantiles denote low-frequency \textit{tail} and higher quantiles denote \textit{head} items. 
Comparing the percentage gain in P@1 of methods wrt the \stdfine model (Fig. \ref{exp:reco_bins}), 
we find that \oldreg helps on the tail items but suffers a huge drop in P@1 on head items. In comparison, \newreg achieves best of both worlds, with the same gains on tail and significantly lower drop on head items. 
Qualitative results for all methods  for the two queries from Section~\ref{sec:qual} are in Suppl.~\ref{sec:appendix_qual}.

\begin{figure*}
\centering
\begin{tikzpicture}
\begin{groupplot}[group style={group size= 2 by 1},width=0.5\textwidth,height=0.30\textwidth]
\nextgroupplot[ybar=1.0pt, 
tick label style={font=\footnotesize},
symbolic x coords={rowe,\#\#nta,z,\#\#100,non,toxic,sole,\#\#plate,cleaner,kit},
xtick={rowe,\#\#nta,z,\#\#100,non,toxic,sole,\#\#plate,cleaner,kit},
ytick={-0.5,-0.25,0,0.25,0.5},
xtick pos=left,
ytick pos=left,
x tick label style={rotate=45,anchor=east},
scaled y ticks = false,
legend pos = north west, 
ymajorgrids=true,
ylabel = {\footnotesize Importance Scores},
legend style={at={(1,1)},anchor=north east,legend columns=1,font=\small},
title={\small \textit{Rowenta Z100 Non Toxic Soleplate Cleaner Kit}},
bar width=3pt]
\addplot+[red, fill=red, postaction={pattern=north west lines}] coordinates {(rowe, 0.073) (\#\#nta, 0.060) (z, 0.060) (\#\#100, 0.027) (non, 0.031) (toxic, 0.075) (sole, 0.019) (\#\#plate, 0.043) (cleaner, 0.031) (kit, 0.034)  };
\addplot+[black!40!white, fill=black!40!white, postaction={pattern=dots}] coordinates {(rowe, 0.421) (\#\#nta, 0.109) (z, 0.039) (\#\#100, 0.049) (non, 0.018) (toxic, 0.025) (sole, 0.014) (\#\#plate, 0.051) (cleaner, 0.084) (kit, 0.010)};
\addplot+[blue, pattern color=blue!40!white, postaction={pattern=crosshatch}] coordinates {(rowe, 0.173) (\#\#nta, 0.056) (z, 0.029) (\#\#100, 0.024) (non, 0.021) (toxic, 0.035) (sole, 0.018) (\#\#plate, 0.077) (cleaner, 0.049) (kit, 0.016)  };

\legend{\base, \stdfine, \newreg};
\coordinate (c1) at (rel axis cs:0,1);

\nextgroupplot[ybar=1.0pt, 
tick label style={font=\footnotesize},
symbolic x coords={sole,\#\#us,air,os,\#\#ci,\#\#lla,\#\#ting,reflective,heat,\#\#er},
xtick={sole,\#\#us,air,os,\#\#ci,\#\#lla,\#\#ting,reflective,heat,\#\#er},
ytick={-0.5,-0.25,0,0.25,0.5},
xtick pos=left,
ytick pos=left,
x tick label style={rotate=45,anchor=east},
scaled y ticks = false,
legend pos = north west, 
ymajorgrids=true,
legend style={at={($(0,0)+(1cm,1cm)$)},legend columns=1,fill=none,draw=black,anchor=center,align=left,legend cell align=left,font=\small},
title={\small \textit{Soleus Air Oscillating Reflective Heater}},
bar width=3pt]
\addplot+[red, fill=red, postaction={pattern=north west lines}] coordinates {(sole, 0.120) (\#\#us, 0.094) (air, 0.082) (os, 0.049) (\#\#ci, 0.049) (\#\#lla, 0.063) (\#\#ting, 0.032) (reflective, 0.091) (heat, 0.073) (\#\#er, 0.058)  };
\addplot+[black!40!white, fill=black!40!white, postaction={pattern=dots}] coordinates {(sole, 0.10) (\#\#us, 0.08) (air, 0.083) (os, 0.016) (\#\#ci, 0.021) (\#\#lla, 0.074) (\#\#ting, 0.064) (reflective, 0.585) (heat, 0.135) (\#\#er, 0.107)  };
\addplot+[blue, pattern color=blue!40!white, postaction={pattern=crosshatch}] coordinates {(sole, 0.14) (\#\#us, 0.102) (air, 0.106) (os, 0.026) (\#\#ci, 0.033) (\#\#lla, 0.052) (\#\#ting, 0.064) (reflective, 0.27) (heat, 0.09) (\#\#er, 0.04)  };

\coordinate (c2) at (rel axis cs:1,1);

\end{groupplot}

\coordinate (c3) at ($(c2)$);

\end{tikzpicture}
\vspace*{-10pt}

\caption{\label{exp:both_examples_itv}Token-wise importance scores (Eqn \ref{eqn:importance}) for two example queries 
using the \base, \stdfine and \newreg. \newreg is able to normalise the importance scores for \textit{"rowe"} and \textit{"reflective"} tokens (as in Fig. \ref{exp:both_examples}) to be closer to \base model.}

\end{figure*}
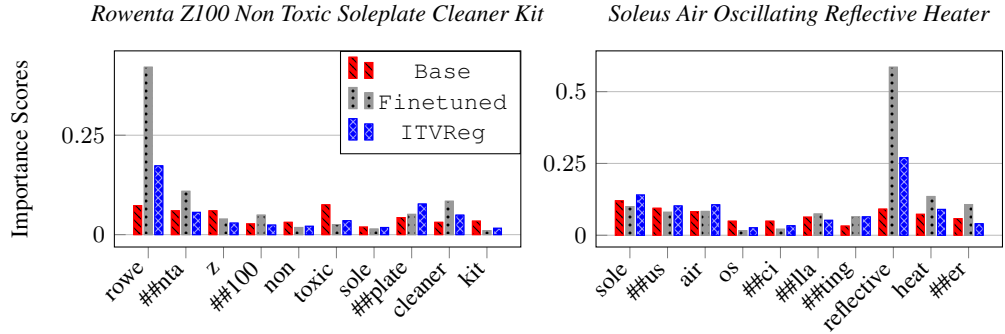

\paragraph{Importance Scores} We also evaluate importance scores for the two qualitative examples presented in Fig. ~\ref{exp:both_examples} after \newreg regularisation in Fig ~\ref{exp:both_examples_itv}. The importance scores for the problematic tokens \textit{rowe} and \textit{reflective} are now in a reasonable range of value guided by the \base model

\subsection{Question recommendation}

\paragraph{Datasets.} 
We use the setup of \cite{shah2018adversarial} for our experiments. The \textit{AskUbuntu} and \textit{SuperUser} data comes from Stack Exchange, a family of technical community support forums. Both datasets contain around one million pairs of sentences, labeled either zero or one denoting a negative and a positive pair. 
Simulating a cold-start scenario with limited items, we present main results on 10\% subsample of these datasets. 
Results on the complete datasets are in Suppl~\ref{sec:complete_results}.
To study generalization under an extreme distribution shift, we additionally evaluate on a non-technical forum,   \textit{Quora-Question Pairs} dataset \cite{wang2018glue}. 

\paragraph{Fine-tuning.} We adopt the same modeling procedure as \cite{thakur2020augmented}, training with a MSE loss between the ground truth and the predicted relevance score.  
We initialize the models using two different \base models: NLI-DistilBERT
~\cite{sanh2019distilbert} 
MSMARCO-DistilBERT-v4
~\cite{sanh2019distilbert} 
the latter having a substantially higher accuracy on our task. For more details on training, refer Supp~\ref{subsec:hyperparameter_question}.

\paragraph{Evaluation metric.} We use AUC(0.05) as the evaluation metric as in  \cite{shah2018adversarial}. AUC(0.05) is the area under the ROC curve when the false positive rate ($fpr$) ranges from 0 to 0.05.

\paragraph{Results.}
Table~\ref{exp:main_table} shows the IID and OOD performance of models trained on the two forum datasets. Let us first consider AUC on evaluation within the technical forums and  using the less accurate NLI-DistilBERT \base model. On both IID and OOD evaluation, \newreg obtains the highest AUC for both datasets. On OOD evaluation, \maskreg and \newaug have the second-best AUC for SuperUser and AskUbuntu evaluation sets respectively.    \oldreg has comparatively lower AUC because the base model is not accurate: \newreg obtains $>10$ points higher AUC on both OOD datasets. 

Using the more accurate MSMARCO-DistilBERT-v4, accuracy of both \oldreg and \newreg increases. P@1 on OOD data of \oldreg is comparable to \newreg on \textit{AskUbuntu} evaluation and 1 point higher than \newreg on \textit{SuperUser} evaluation. In comparison,  \maskreg and \simcse fail to  utilise the better \base model. 
When evaluated on an extreme distribution shift (\textit{Quora}), \base model performs the best on OOD evaluation 
(Table~\ref{exp:main_table})
, followed by \oldreg. Finetuning on \textit{AskUbuntu} or \textit{SuperUser}  adds no  information. 

\begin{table*}
\small
\centering
\resizebox{\textwidth}{!}{%
\begin{tabular}{ll|lll|lll}
\hline
\textbf{\multirow{2}{*}{Base Model}} & \textbf{\multirow{2}{*}{Method}} & \multicolumn{3}{c|}{Train : AskUbuntu} & \multicolumn{3}{c}{Train : Superuser} \\
& & \textbf{AskUbuntu} & \textbf{Superuser} & \textbf{Quora} & \textbf{AskUbuntu} & \textbf{Superuser} & \textbf{Quora} \\
\hline

\multirow{7}{*}{\small{NLI}} & \base & 31.80 & 48.50 & \textbf{14.00} & 31.80 & 48.50 & \textbf{14.00} \\\cline{2-8}
& \stdfine& 65.68 $\pm$ 0.54& 72.70 $\pm$ 0.73 & 12.74 $\pm$ 0.43 & \textbf{47.98 $\pm$ 6.63}& 75.70 $\pm$ 1.49 & 12.82 $\pm$ 0.49\\\cline{2-8} 
& \maskreg & \textbf{71.51 $\pm$ 1.16} & 77.55 $\pm$ 0.57 & 12.44 $\pm$ 0.48& 33.83 $\pm$ 3.07& 81.97 $\pm$ 0.33 & 13.13 $\pm$ 0.17\\\cline{2-8}
& \simcse & 67.31 $\pm$ 1.28& 73.46 $\pm$ 2.24& 12.87 $\pm$ 0.36& 45.03 $\pm$ 4.22& 76.80 $\pm$ 0.68 & 13.41 $\pm$ 0.20\\\cline{2-8}
& \oldreg& 56.86 $\pm$ 0.69& 66.28 $\pm$ 0.66 & \textbf{14.09 $\pm$ 0.18} & 36.52 $\pm$ 1.53& 71.11 $\pm$ 0.86 & \textbf{14.44 $\pm$ 0.28} \\\cline{2-8}
& \newaug & 70.10 $\pm$ 0.80 & 76.98 $\pm$ 0.33 & 13.74 $\pm$ 0.12 & 45.86 $\pm$ 4.10 & 81.34 $\pm$ 0.11 & 14.22 $\pm$ 0.45\\\cline{2-8}
& \newreg & \textbf{71.24 $\pm$ 0.62} & \textbf{78.89 $\pm$ 0.59} & 13.57 $\pm$ 0.32 & \textbf{47.00 $\pm$ 1.58} & \textbf{83.40 $\pm$ 0.54} & 14.67 $\pm$ 0.18\\\cline{1-8}    

\multirow{7}{*}{\small{MSMARCO}} & \base & 54.01 &  80.73 &  \textbf{18.39} & 54.01 &  80.73 & \textbf{18.39}\\\cline{2-8}  
& \stdfine & 68.11 $\pm$ 1.09& 75.33 $\pm$ 0.74 & 14.05 $\pm$ 0.19 & 59.27 $\pm$ 2.16& 79.28 $\pm$ 0.94 & 14.35 $\pm$ 0.20 \\\cline{2-8}
& \maskreg& 72.59 $\pm$ 0.41& 79.00 $\pm$ 0.08 & 12.63 $\pm$ 0.36 & 32.05 $\pm$ 6.53& 83.53 $\pm$ 0.53 & 13.65 $\pm$ 0.20 \\\cline{2-8}
& \simcse & 70.07 $\pm$ 0.32& 76.54 $\pm$ 1.43 & 14.98 $\pm$ 0.08 & 48.97 $\pm$ 3.28& 80.51 $\pm$ 0.70 & 14.70 $\pm$ 0.05\\\cline{2-8}
& \oldreg& 73.04 $\pm$ 0.77& \textbf{84.09 $\pm$ 0.28}& 17.16 $\pm$ 0.13 & \textbf{60.75 $\pm$ 0.84} & \textbf{86.24 $\pm$ 0.13} & 17.64 $\pm$ 0.25 \\\cline{2-8}
& \newaug & 73.17 $\pm$ 0.46& 80.53 $\pm$ 0.23 & 15.36 $\pm$ 0.48 & 54.85 $\pm$ 3.94& 84.29 $\pm$ 0.35 & 15.64 $\pm$ 0.32 \\\cline{2-8}
& \newreg & \textbf{74.64 $\pm$ 0.56} & 82.86 $\pm$ 0.65  & 16.11 $\pm$ 0.36 & \textbf{60.07 $\pm$ 2.42} & \textbf{86.24 $\pm$ 0.24} & 16.39 $\pm$ 0.31\\\cline{1-8}    

\hline
\end{tabular}
}
\caption{\label{exp:main_table}
AUC (0.05) using two different base models, NLI-DistilBERT-Base and MSMARCO-DistilBERT-Base. First three columns correspond to training on \textit{AskUbuntu} and the  last three  training on \textit{SuperUser}. When evaluated on technical forums with a weaker base model (NLI), \newreg obtains the best AUC on both IID and OOD evaluation. }
\end{table*}

\section{Discussion and Conclusion}
We identified limitations of text-matching recommender systems on OOD generalization, showing that state-of-the-art models often perform worse than the base model on which they were finetuned. We proposed two regularizers using the base model, \newreg and \oldreg.  Compared to other methods, \newreg obtains high IID accuracy and OOD accuracy, thus exploiting the best of the \base model and the training data. However, under extreme distribution shifts where IID training data is not useful, \oldreg or the \base model is more suitable.

\section*{Limitations}
We studied a specific setting of short text-matching recommender systems and our results may not apply to other kinds of text-based recommender systems that utilize longer descriptions or other metadata. Another limitation is that all our experiments use offline data collected from recommender systems. It will be useful to study the performance of proposed methods when deployed on a real-world system that faces distribution shifts. Finally, the potential of our method for improving OOD generalization depends on the quality of the base model. 

\bibliography{custom}

\begin{thebibliography}{42}
\expandafter\ifx\csname natexlab\endcsname\relax\def\natexlab#1{#1}\fi

\bibitem[{Abdollahpouri et~al.(2019)Abdollahpouri, Mansoury, Burke, and
  Mobasher}]{abdollahpouri2019unfairness}
Himan Abdollahpouri, Masoud Mansoury, Robin Burke, and Bamshad Mobasher. 2019.
\newblock The unfairness of popularity bias in recommendation.
\newblock \emph{arXiv preprint arXiv:1907.13286}.

\bibitem[{Adomavicius and Tuzhilin(2005)}]{adomavicius2005toward}
Gediminas Adomavicius and Alexander Tuzhilin. 2005.
\newblock Toward the next generation of recommender systems: A survey of the
  state-of-the-art and possible extensions.
\newblock \emph{IEEE transactions on knowledge and data engineering},
  17(6):734--749.

\bibitem[{Bastings and Filippova(2020)}]{bastings2020elephant}
Jasmijn Bastings and Katja Filippova. 2020.
\newblock The elephant in the interpretability room: Why use attention as
  explanation when we have saliency methods?
\newblock \emph{arXiv preprint arXiv:2010.05607}.

\bibitem[{Bhatia et~al.(2016)Bhatia, Dahiya, Jain, Kar, Mittal, Prabhu, and
  Varma}]{Bhatia16}
K.~Bhatia, K.~Dahiya, H.~Jain, P.~Kar, A.~Mittal, Y.~Prabhu, and M.~Varma.
  2016.
\newblock \href {http://manikvarma.org/downloads/XC/XMLRepository.html} {The
  extreme classification repository: Multi-label datasets and code}.

\bibitem[{Broder et~al.(2008)Broder, Ciccolo, Fontoura, Gabrilovich,
  Josifovski, and Riedel}]{broder2008search}
Andrei~Z Broder, Peter Ciccolo, Marcus Fontoura, Evgeniy Gabrilovich, Vanja
  Josifovski, and Lance Riedel. 2008.
\newblock Search advertising using web relevance feedback.
\newblock In \emph{Proceedings of the 17th ACM conference on information and
  knowledge management}, pages 1013--1022.

\bibitem[{Cha et~al.(2022)Cha, Lee, Park, and Chun}]{cha2022domain}
Junbum Cha, Kyungjae Lee, Sungrae Park, and Sanghyuk Chun. 2022.
\newblock Domain generalization by mutual-information regularization with
  pre-trained models.
\newblock \emph{arXiv preprint arXiv:2203.10789}.

\bibitem[{Chang et~al.(2020)Chang, Yu, Chang, Yang, and Kumar}]{chang2020pre}
Wei-Cheng Chang, Felix~X Yu, Yin-Wen Chang, Yiming Yang, and Sanjiv Kumar.
  2020.
\newblock Pre-training tasks for embedding-based large-scale retrieval.
\newblock \emph{arXiv preprint arXiv:2002.03932}.

\bibitem[{Dahiya et~al.(2021)Dahiya, Agarwal, Saini, Gururaj, Jiao, Singh,
  Agarwal, Kar, and Varma}]{dahiya2021siamesexml}
Kunal Dahiya, Ananye Agarwal, Deepak Saini, K~Gururaj, Jian Jiao, Amit Singh,
  Sumeet Agarwal, Purushottam Kar, and Manik Varma. 2021.
\newblock Siamesexml: Siamese networks meet extreme classifiers with 100m
  labels.
\newblock In \emph{ICML}.

\bibitem[{Devlin et~al.(2018)Devlin, Chang, Lee, and
  Toutanova}]{devlin2018bert}
Jacob Devlin, Ming-Wei Chang, Kenton Lee, and Kristina Toutanova. 2018.
\newblock Bert: Pre-training of deep bidirectional transformers for language
  understanding.
\newblock \emph{arXiv preprint arXiv:1810.04805}.

\bibitem[{Gao and Callan(2021)}]{gao2021condenser}
Luyu Gao and Jamie Callan. 2021.
\newblock Condenser: a pre-training architecture for dense retrieval.
\newblock \emph{arXiv preprint arXiv:2104.08253}.

\bibitem[{Gao et~al.(2021)Gao, Yao, and Chen}]{gao2021simcse}
Tianyu Gao, Xingcheng Yao, and Danqi Chen. 2021.
\newblock Simcse: Simple contrastive learning of sentence embeddings.
\newblock \emph{arXiv preprint arXiv:2104.08821}.

\bibitem[{He et~al.(2022)He, Wang, Cui, Zou, Zhang, Cui, and
  Jiang}]{he2022causpref}
Yue He, Zimu Wang, Peng Cui, Hao Zou, Yafeng Zhang, Qiang Cui, and Yong Jiang.
  2022.
\newblock Causpref: Causal preference learning for out-of-distribution
  recommendation.
\newblock \emph{arXiv preprint arXiv:2202.03984}.

\bibitem[{Hendrycks et~al.(2020)Hendrycks, Liu, Wallace, Dziedzic, Krishnan,
  and Song}]{hendrycks2020pretrained}
Dan Hendrycks, Xiaoyuan Liu, Eric Wallace, Adam Dziedzic, Rishabh Krishnan, and
  Dawn Song. 2020.
\newblock Pretrained transformers improve out-of-distribution robustness.
\newblock \emph{arXiv preprint arXiv:2004.06100}.

\bibitem[{Hoang et~al.(2022)Hoang, Deoras, Zhao, Li, and
  Karypis}]{hoang2022learning}
Trong~Nghia Hoang, Anoop Deoras, Tong Zhao, Jin Li, and George Karypis. 2022.
\newblock Learning personalized item-to-item recommendation metric via implicit
  feedback.
\newblock \emph{arXiv preprint arXiv:2203.12598}.

\bibitem[{Karpukhin et~al.(2020)Karpukhin, O{\u{g}}uz, Min, Lewis, Wu, Edunov,
  Chen, and Yih}]{karpukhin2020dense}
Vladimir Karpukhin, Barlas O{\u{g}}uz, Sewon Min, Patrick Lewis, Ledell Wu,
  Sergey Edunov, Danqi Chen, and Wen-tau Yih. 2020.
\newblock Dense passage retrieval for open-domain question answering.
\newblock \emph{arXiv preprint arXiv:2004.04906}.

\bibitem[{Kumar et~al.(2022)Kumar, Raghunathan, Jones, Ma, and
  Liang}]{kumar2022fine}
Ananya Kumar, Aditi Raghunathan, Robbie Jones, Tengyu Ma, and Percy Liang.
  2022.
\newblock Fine-tuning can distort pretrained features and underperform
  out-of-distribution.
\newblock \emph{arXiv preprint arXiv:2202.10054}.

\bibitem[{Lee et~al.(2019)Lee, Chang, and Toutanova}]{lee2019latent}
Kenton Lee, Ming-Wei Chang, and Kristina Toutanova. 2019.
\newblock Latent retrieval for weakly supervised open domain question
  answering.
\newblock \emph{arXiv preprint arXiv:1906.00300}.

\bibitem[{Lee et~al.(2022)Lee, Yao, and Finn}]{lee2022diversify}
Yoonho Lee, Huaxiu Yao, and Chelsea Finn. 2022.
\newblock Diversify and disambiguate: Learning from underspecified data.
\newblock \emph{arXiv preprint arXiv:2202.03418}.

\bibitem[{Linden et~al.(2003)Linden, Smith, and York}]{linden2003amazon}
Greg Linden, Brent Smith, and Jeremy York. 2003.
\newblock Amazon. com recommendations: Item-to-item collaborative filtering.
\newblock \emph{IEEE Internet computing}, 7(1):76--80.

\bibitem[{Mahajan et~al.(2021)Mahajan, Tople, and Sharma}]{mahajan2021domain}
Divyat Mahajan, Shruti Tople, and Amit Sharma. 2021.
\newblock Domain generalization using causal matching.
\newblock In \emph{International Conference on Machine Learning}, pages
  7313--7324. PMLR.

\bibitem[{Makar et~al.(2022)Makar, Packer, Moldovan, Blalock, Halpern, and
  D’Amour}]{makar2022causally}
Maggie Makar, Ben Packer, Dan Moldovan, Davis Blalock, Yoni Halpern, and
  Alexander D’Amour. 2022.
\newblock Causally motivated shortcut removal using auxiliary labels.
\newblock In \emph{International Conference on Artificial Intelligence and
  Statistics}, pages 739--766. PMLR.

\bibitem[{Mittal et~al.(2021)Mittal, Sachdeva, Agrawal, Agarwal, Kar, and
  Varma}]{mittal2021eclare}
Anshul Mittal, Noveen Sachdeva, Sheshansh Agrawal, Sumeet Agarwal, Purushottam
  Kar, and Manik Varma. 2021.
\newblock Eclare: Extreme classification with label graph correlations.
\newblock In \emph{Proceedings of the Web Conference 2021}, pages 3721--3732.

\bibitem[{Reimers and Gurevych(2019)}]{reimers2019sentence}
Nils Reimers and Iryna Gurevych. 2019.
\newblock Sentence-bert: Sentence embeddings using siamese bert-networks.
\newblock \emph{arXiv preprint arXiv:1908.10084}.

\bibitem[{Ribeiro et~al.(2020)Ribeiro, Wu, Guestrin, and
  Singh}]{ribeiro2020beyond}
Marco~Tulio Ribeiro, Tongshuang Wu, Carlos Guestrin, and Sameer Singh. 2020.
\newblock Beyond accuracy: Behavioral testing of nlp models with checklist.
\newblock \emph{arXiv preprint arXiv:2005.04118}.

\bibitem[{Sagawa et~al.(2019)Sagawa, Koh, Hashimoto, and
  Liang}]{sagawa2019distributionally}
Shiori Sagawa, Pang~Wei Koh, Tatsunori~B Hashimoto, and Percy Liang. 2019.
\newblock Distributionally robust neural networks for group shifts: On the
  importance of regularization for worst-case generalization.
\newblock \emph{arXiv preprint arXiv:1911.08731}.

\bibitem[{Saini et~al.(2021)Saini, Jain, Dave, Jiao, Singh, Zhang, and
  Varma}]{saini2021galaxc}
Deepak Saini, Arnav~Kumar Jain, Kushal Dave, Jian Jiao, Amit Singh, Ruofei
  Zhang, and Manik Varma. 2021.
\newblock Galaxc: Graph neural networks with labelwise attention for extreme
  classification.
\newblock In \emph{Proceedings of the Web Conference 2021}, pages 3733--3744.

\bibitem[{Sanh et~al.(2019)Sanh, Debut, Chaumond, and
  Wolf}]{sanh2019distilbert}
Victor Sanh, Lysandre Debut, Julien Chaumond, and Thomas Wolf. 2019.
\newblock Distilbert, a distilled version of bert: smaller, faster, cheaper and
  lighter.
\newblock \emph{arXiv preprint arXiv:1910.01108}.

\bibitem[{Schnabel and Bennett(2020)}]{schnabel2020debiasing}
Tobias Schnabel and Paul~N Bennett. 2020.
\newblock Debiasing item-to-item recommendations with small annotated datasets.
\newblock In \emph{Fourteenth ACM Conference on Recommender Systems}, pages
  73--81.

\bibitem[{Shah et~al.(2018)Shah, Lei, Moschitti, Romeo, and
  Nakov}]{shah2018adversarial}
Darsh~J Shah, Tao Lei, Alessandro Moschitti, Salvatore Romeo, and Preslav
  Nakov. 2018.
\newblock Adversarial domain adaptation for duplicate question detection.
\newblock \emph{arXiv preprint arXiv:1809.02255}.

\bibitem[{Thakur et~al.(2020)Thakur, Reimers, Daxenberger, and
  Gurevych}]{thakur2020augmented}
Nandan Thakur, Nils Reimers, Johannes Daxenberger, and Iryna Gurevych. 2020.
\newblock Augmented sbert: Data augmentation method for improving bi-encoders
  for pairwise sentence scoring tasks.
\newblock \emph{arXiv preprint arXiv:2010.08240}.

\bibitem[{Wang et~al.(2018)Wang, Singh, Michael, Hill, Levy, and
  Bowman}]{wang2018glue}
Alex Wang, Amanpreet Singh, Julian Michael, Felix Hill, Omer Levy, and Samuel~R
  Bowman. 2018.
\newblock Glue: A multi-task benchmark and analysis platform for natural
  language understanding.
\newblock \emph{arXiv preprint arXiv:1804.07461}.

\bibitem[{Wang et~al.(2022)Wang, Lan, Liu, Ouyang, Qin, Lu, Chen, Zeng, and
  Yu}]{wang2022generalizing}
Jindong Wang, Cuiling Lan, Chang Liu, Yidong Ouyang, Tao Qin, Wang Lu, Yiqiang
  Chen, Wenjun Zeng, and Philip Yu. 2022.
\newblock Generalizing to unseen domains: A survey on domain generalization.
\newblock \emph{IEEE Transactions on Knowledge and Data Engineering}.

\bibitem[{Wortsman et~al.(2021)Wortsman, Ilharco, Li, Kim, Hajishirzi, Farhadi,
  Namkoong, and Schmidt}]{wortsman2021robust}
Mitchell Wortsman, Gabriel Ilharco, Mike Li, Jong~Wook Kim, Hannaneh
  Hajishirzi, Ali Farhadi, Hongseok Namkoong, and Ludwig Schmidt. 2021.
\newblock Robust fine-tuning of zero-shot models.
\newblock \emph{arXiv preprint arXiv:2109.01903}.

\bibitem[{Wu et~al.(2020)Wu, Wang, Gu, Khabsa, Sun, and Ma}]{wu2020clear}
Zhuofeng Wu, Sinong Wang, Jiatao Gu, Madian Khabsa, Fei Sun, and Hao Ma. 2020.
\newblock Clear: Contrastive learning for sentence representation.
\newblock \emph{arXiv preprint arXiv:2012.15466}.

\bibitem[{Xie et~al.(2020)Xie, Sun, Liu, Wu, Gao, Ding, and
  Cui}]{xie2020contrastiveaugs}
Xu~Xie, Fei Sun, Zhaoyang Liu, Shiwen Wu, Jinyang Gao, Bolin Ding, and Bin Cui.
  2020.
\newblock Contrastive learning for sequential recommendation.
\newblock \emph{arXiv preprint arXiv:2010.14395}.

\bibitem[{Xiong et~al.(2020)Xiong, Xiong, Li, Tang, Liu, Bennett, Ahmed, and
  Overwijk}]{xiong2020approximate}
Lee Xiong, Chenyan Xiong, Ye~Li, Kwok-Fung Tang, Jialin Liu, Paul~N Bennett,
  Junaid Ahmed, and Arnold Overwijk. 2020.
\newblock Approximate nearest neighbor negative contrastive learning for dense
  text retrieval.
\newblock In \emph{International Conference on Learning Representations}.

\bibitem[{Xiong et~al.(2021)Xiong, Xiong, Li, Tang, Liu, Bennett, Ahmed, and
  Overwijk}]{lee2021ance}
Lee Xiong, Chenyan Xiong, Ye~Li, Kwok{-}Fung Tang, Jialin Liu, Paul~N. Bennett,
  Junaid Ahmed, and Arnold Overwijk. 2021.
\newblock Approximate nearest neighbor negative contrastive learning for dense
  text retrieval.
\newblock In \emph{ICLR}.

\bibitem[{Yao et~al.(2022)Yao, Wang, Li, Zhang, Liang, Zou, and
  Finn}]{yao2022improving}
Huaxiu Yao, Yu~Wang, Sai Li, Linjun Zhang, Weixin Liang, James Zou, and Chelsea
  Finn. 2022.
\newblock Improving out-of-distribution robustness via selective augmentation.
\newblock \emph{arXiv preprint arXiv:2201.00299}.

\bibitem[{Yu et~al.(2021)Yu, Yin, Xia, Chen, Cui, and
  Nguyen}]{yugraphaugcritique2021}
Junliang Yu, Hongzhi Yin, Xin Xia, Tong Chen, Lizhen Cui, and Quoc Viet~Hung
  Nguyen. 2021.
\newblock \href {https://doi.org/10.48550/ARXIV.2112.08679} {Are graph
  augmentations necessary? simple graph contrastive learning for
  recommendation}.

\bibitem[{Yu et~al.(2022)Yu, Yin, Xia, Chen, Li, and
  Huang}]{yu2022selfsupervisedsurvey}
Junliang Yu, Hongzhi Yin, Xin Xia, Tong Chen, Jundong Li, and Zi~Huang. 2022.
\newblock Self-supervised learning for recommender systems: A survey.
\newblock \emph{arXiv preprint arXiv:2203.15876}.

\bibitem[{Zhou et~al.(2021{\natexlab{a}})Zhou, Ma, Zhang, Zhou, and
  Yang}]{zhou2021contrastive}
Chang Zhou, Jianxin Ma, Jianwei Zhang, Jingren Zhou, and Hongxia Yang.
  2021{\natexlab{a}}.
\newblock Contrastive learning for debiased candidate generation in large-scale
  recommender systems.
\newblock In \emph{Proceedings of the 27th ACM SIGKDD Conference on Knowledge
  Discovery \& Data Mining}, pages 3985--3995.

\bibitem[{Zhou et~al.(2021{\natexlab{b}})Zhou, Liu, Qiao, Xiang, and
  Loy}]{zhou2021domain}
Kaiyang Zhou, Ziwei Liu, Yu~Qiao, Tao Xiang, and Chen~Change Loy.
  2021{\natexlab{b}}.
\newblock Domain generalization in vision: A survey.
\newblock \emph{arXiv preprint arXiv:2103.02503}.

\end{thebibliography}
\bibliographystyle{acl_natbib}

\appendix



\section{Dataset Details }
\paragraph{Quora, SuperUser, AskUbuntu} The ratio of positives to negatives in \textit{SuperUser, AskUbuntu} is 1:100, while in \textit{Quora} is 4:7.

\paragraph{Amazon131K} In \textit{Amazon131K} we have categorical information for 99K labels. The categorical shift experiments are conducted on this filtered data.

\section{Hyper-parameter Tuning }
\label{sec:hyperparameter_tuning}
For \simcse the only choice of hyperparameter is the $\lambda$ regularisation coefficient. We search over 3 values of $\lambda$ i.e. 0.01,0.1,1.0 and select 0.1 as the best value. For other methods also we fix $\lambda$ as 0.1. See Table ~\ref{tab:simcse_tune}
\begin{table}[h]
\centering
\begin{tabular}{llll}
\hline
\textbf{\small Method} & \textbf{\small Quora} & \textbf{\small AskUbuntu} & \textbf{\small Superuser} \\
\hline

SimCSE 0.01 & 54.98 & 73.70 & 84.91  \\\cline{1-4}  
SimCSE 0.1 & 55.40 & 74.34 & 86.38 \\\cline{1-4}
SimCSE 1.0 & 54.89 & 74.07 & 85.41  \\\cline{1-4} 

\hline
\end{tabular}
\caption{\label{tab:simcse_tune}
\simcse IID numbers finetuned with base model as MSMARCO-DistilBERT-v4 on complete \textit{Quora},\textit{AskUbuntu},\textit{SuperUser} datasets. This shows that 0.1 is the optimal hyperparameter for \simcse}
\end{table}

For \maskreg and \newreg another hyperparameter choice is the masking fraction of the input. We search over 2 choices of masking namely masking 50\% of input, or 15\% of input. We find that 15\% works best for \maskreg while 50\% gives best results for \newreg. We use these parameters for all experiments. For \maskreg results are reported in Table ~\ref{tab:masking_tune}, while for \newreg results can be seen in Table ~\ref{exp:quora_complete},\ref{exp:superuser_complete},\ref{exp:ubuntu_complete}.
\begin{table*}
\small
\centering
\begin{tabular}{llllll}
\hline
\textbf{\multirow{2}{*}{Base Model}} & \textbf{\multirow{2}{*}{Method}} & \multicolumn{2}{c}{Train : AskUbuntu} & \multicolumn{2}{c}{Train : Superuser} \\
& & \textbf{AskUbuntu} & \textbf{Superuser} & \textbf{AskUbuntu} & \textbf{Superuser}\\
\hline

\multirow{3}{*}{\small{NLI}} & \base & 31.80 & 48.50 & 31.80 & 48.50 \\\cline{2-6}
& \maskreg (0.15) & 71.51 $\pm$ 1.16& 77.55 $\pm$ 0.57& 33.83 $\pm$ 3.07& 81.97 $\pm$ 0.33\\\cline{2-6}
& \maskreg (0.50) & 64.57 $\pm$ 0.68 & 69.87 $\pm$ 0.36& 19.85 $\pm$ 4.14& 77.17 $\pm$ 0.98\\\cline{1-6}

\multirow{3}{*}{\small{MSMARCO}} & \base & 54.01 &  80.73 & 54.01 &  80.73 \\\cline{2-6}  
& \maskreg (0.15) & 72.59 $\pm$ 0.41& 79.00 $\pm$ 0.08& 32.05 $\pm$ 6.53& 83.53 $\pm$ 0.53\\\cline{2-6}
& \maskreg (0.50) & 64.42 $\pm$ 1.44 & 71.56 $\pm$ 0.56 & 20.36 $\pm$ 5.69& 77.64 $\pm$ 0.94\\\cline{1-6}

\hline
\end{tabular}
\caption{\label{tab:masking_tune}
AUC (0.05) for \maskreg with 15\% and 50\% input token masking. 15\% is the optimal masking for \maskreg. The setup is same as in Table \ref{exp:main_table}.}
\end{table*}

We also try running with lower learning rates but higher learning rate gives better numbers and hence we work with 1e-4. See Table ~\ref{tab:gridsearch} for reference. 
\begin{table}
\centering
\begin{tabular}{lllll}
\hline
\textbf{\small Learning Rate} & \textbf{\small 2 Epochs} & \textbf{\small 4 Epochs} & \textbf{\small 10 Epochs}\\
\hline

1e-5 & 39.15 & 45.05 & 53.40 \\\cline{1-4}  
5e-5 & 50.67 & 54.87 & 57.20 \\\cline{1-4}
1e-4 & 52.14 & 55.54 & 57.80 \\\cline{1-4} 

\hline
\end{tabular}
\caption{\label{tab:gridsearch}
AUC(0.05) on IID complete \textit{Quora} with \stdfine method on NLI-DistilBERT-Base model. Higher learning rate and more epochs help in training. For computational efficiency we hence take 4 epochs for complete setting and 20 epochs for 10\% \textit{Quora} subset.}
\end{table}

Since we deal with short sentences, the max token length is fixed at 32 which allows for bigger batch sizes (900 for Amazon Titles, 250 for Question Recommendation). 

For \newreg and \oldreg we report numbers for various values of hyperparameter $\lambda$ in Table ~\ref{exp:lambda_sensitivity}. We can see that for higher values of $\lambda$ \oldreg behaves more like the \base model but the numbers for both \newreg and \oldreg are stable across this wide range of $\lambda$.
\begin{table*}[ht]
\small
\centering
\begin{tabular}{ll|lll}
\hline
\textbf{\multirow{2}{*}{Base Model}} & \textbf{\multirow{2}{*}{Method}} & \multicolumn{3}{c
}{Train : Quora}  \\
& & \textbf{Quora} & \textbf{AskUbuntu} & \textbf{Superuser} \\
\hline

\multirow{6}{*}{\small{MSMARCO-DistilBERT}} & \base & 0.184 $\pm$ 0.000 & 0.540 $\pm$ 0.000 & 0.807 $\pm$ 0.000 \\\cline{2-5}

& \newreg 0.01 & 0.550 $\pm$ 0.031 & 0.165 $\pm$ 0.023 & 0.293 $\pm$ 0.018 \\\cline{2-5}
& \newreg 0.1 & 0.577 $\pm$ 0.002 & 0.175 $\pm$ 0.001 & 0.286 $\pm$ 0.011 \\\cline{2-5}
& \newreg 1.0 & 0.577 $\pm$ 0.002 & 0.175 $\pm$ 0.001 & 0.286 $\pm$ 0.011 \\\cline{2-5}
& \oldreg 0.01 & 0.570 $\pm$ 0.006 & 0.176 $\pm$ 0.023 & 0.278 $\pm$ 0.002 \\\cline{2-5}
& \oldreg 0.1 & 0.558 $\pm$ 0.005 & 0.297 $\pm$ 0.022 & 0.438 $\pm$ 0.006 \\\cline{2-5}
& \oldreg 1.0 & 0.491 $\pm$ 0.015 & 0.456 $\pm$ 0.001 & 0.646 $\pm$ 0.008 \\\cline{2-5}

\hline
\end{tabular}
\caption{\label{exp:lambda_sensitivity}
AUC (0.05) for different values of the hyperparameter $\lambda$. We can see that both \newreg and \oldreg behave well in these reasonable range of $\lambda$. \oldreg on higher values of hyperparameter imitates the \base model while \newreg still learns useful information from the data}
\end{table*}

\section{Training Details}

\subsection{AmazonTitle Recommendations Training Details} \label{subsec:hyperparameter_amazon}
We follow the modeling procedure used in a recent state-of-the-art work \cite{dahiya2021siamesexml}. For training, We use contrastive loss with hard negative mining. we initialise the model as MSMARCO-DistilBERT model, and fine-tune it for 200 epochs with a batch size of 900. The only difference from \cite{dahiya2021siamesexml}'s setup is that we train only till 200 epochs instead of convergence (300-400 epochs) due to computational constraints. We use Adam optimizer with learning rate of 1e-4.

\subsection{Question Recommendations Training Details} \label{subsec:hyperparameter_question}
We train all the models for 5 epochs with a batch size of 250. We use learning rate of 1e-4 to fine-tune the model. We train the model with a MSE loss between the ground truth and the predicted similarities, as done in \cite{thakur2020augmented}. For \textit{Quora} on the 10\% subset, we train the model for 20 epochs with lr of 1e-4. 

\section{Category-wise P@1 for AmazonCatOOD}
We report category wise numbers OOD numbers for AmazonCatOOD (Table \ref{tab:reco}). All categories have a similar trend. Results can be seen in Table \ref{exp:cat_numbers}.
\label{sec:amzn_cat_numbers}
\begin{table*}
\small
\centering
\resizebox{\textwidth}{!}{%
\begin{tabular}{llllll}
\hline
\textbf{Fine-tuning Method} & \textbf{Automobiles} & \textbf{Kitchen and Dining} & \textbf{Health and Personal Care} & \textbf{Electronics} & \textbf{Tools and Home Imp.} \\
\hline

No Finetune & 32.66 & 30.68 & 30.42 & 32.07 & 31.67 \\\cline{1-6}  
Std Finetune & 31.41 $\pm$ 0.28 & 29.70 $\pm$ 0.27 & 29.17 $\pm$ 0.25 & 30.60 $\pm$ 0.13 & 30.59 $\pm$ 0.25 \\\cline{1-6} 
\oldreg & 34.16 $\pm$ 0.04 & 32.25 $\pm$ 0.16 & 31.75 $\pm$ 0.11& 33.14 $\pm$ 0.07 & 33.18 $\pm$ 0.10 \\\cline{1-6}
\maskreg & 32.16 $\pm$ 0.23 & 30.30 $\pm$ 0.18 & 29.85 $\pm$ 0.24& 31.24 $\pm$ 0.11 & 31.22 $\pm$ 0.23\\\cline{1-6}
SimCSE & 31.45 $\pm$ 0.12 & 29.84 $\pm$ 0.04 & 29.38 $\pm$ 0.06& 30.70 $\pm$ 0.19 & 30.61 $\pm$ 0.16\\\cline{1-6}    
\newreg  & 32.61 $\pm$ 0.02 & 30.76 $\pm$ 0.10 & 30.26 $\pm$ 0.18& 31.73 $\pm$ 0.14 & 31.59 $\pm$ 0.08\\\cline{1-6}

\hline
\end{tabular}
}
\caption{\label{exp:cat_numbers}
Category wise numbers OOD numbers for \textit{AmazonCatOOD} (Table \ref{tab:reco}). All categories have a similar trend as observed in Table \ref{tab:reco}. This shows that our results are agnostic to the choice of categories and hold true for any category in general. 
}
\end{table*}

\section{Evaluation Metrics}
\subsection{Precision@1} \label{sec:pat1}
Given a query, the model outputs a  vector $\hat{R} \in \mathbb{R}^M$, where $M$ is the number of items and $\hat{r}_j$ denotes the similarity between the query and the $j$ th item. We use the $Precision@1$ metric for evaluation, interpreted as the fraction of queries where the top-predicted item is in the query's ground-truth relevant items. Given predicted similarities $\hat{r}$ and ground truth similarities $r$ (where $r_i$ is 1 if its a relevant item, otherwise 0), $P@k$ is defined as $
P@k = \frac{1}{k}\underset{l \in rank_k(\hat{r})}{\sum}r_l
$.

\section{Qualitative Samples}{\label{sec:appendix_qual}}
We give top 5 predicted labels as qualitative samples for "\textit{Soleus Air Oscillating Reflective Heater}" query in Table~\ref{tab:qualitative_reflective} and for query " \textit{Rowenta Z100 Non Toxic Soleplate Cleaner}" in Table~\ref{tab:qualitative_rowe}.

\begin{table*}
\small
\centering
\begin{tabular}{l|l}
\hline Method & Top 5 Predicted Items \\
\hline
\multirow{5}{*}{\base} & Ridgid 31105 24-Inch Aluminum Pipe \textbf{Wrench} \\\cline{2-2}
& Ridgid 31115 48-Inch Aluminum Pipe \textbf{Wrench} \\\cline{2-2}
& Ridgid 31110 36-Inch Aluminum Pipe \textbf{Wrench} \\\cline{2-2}
& Ridgid 31100 18-Inch Aluminum Pipe \textbf{Wrench} \\\cline{2-2}
& Craftsman 9-41796 Ratcheting Ready Bit Screwdriver \\\cline{1-2}

\multirow{5}{*}{\stdfine} & Rowenta ZD100 Non-Toxic Soleplate Cleaner Kit \\\cline{2-2}
& Rowenta DR5015 800 Watt Ultra Steam Brush with Travel Pouch \\\cline{2-2}
& Rowenta(R) Stainless Steel Soleplate Cleaning Kit ZD-110 \\\cline{2-2}
& Rowenta DR6015 Ultrasteam Hand-Held Steam Brush with Travel Pouch, 800-watt \\\cline{2-2}
& Rowenta DR6050 Ultrasteam Hand-Held Steam Brush Dual-Voltage with Travel Pouch, 800-watt \\\cline{1-2}

\multirow{5}{*}{\oldreg } & Rowenta DR6015 Ultrasteam Hand-Held Steam Brush with Travel Pouch, 800-watt \\\cline{2-2}
& Rowenta DW4060 Auto Steam Iron 1700W with Airglide Stainless Steel Soleplate Auto-off Anti-Scale, Blue \\\cline{2-2}
& Rowenta DR5015 800 Watt Ultra Steam Brush with Travel Pouch \\\cline{2-2}
& Rowenta VU2531 Turbo Silence 4-Speed Oscillating Desk Fan, 12-Inch, Bronze \\\cline{2-2}
& Rowenta(R) Stainless Steel Soleplate Cleaning Kit ZD-110 \\\cline{1-2}

\multirow{5}{*}{\maskreg} & Rowenta(R) Stainless Steel Soleplate Cleaning Kit ZD-110 \\\cline{2-2}
& Rowenta ZD100 Non-Toxic Soleplate Cleaner Kit \\\cline{2-2}
& Rowenta DG8430 Pro Precision Steam Station with 400 hole Stainless Steel soleplate 1800 Watt, Purple \\\cline{2-2}
& Rowenta DR5015 800 Watt Ultra Steam Brush with Travel Pouch \\\cline{2-2}
& Rowenta DR6015 Ultrasteam Hand-Held Steam Brush with Travel Pouch, 800-watt \\\cline{1-2}

\multirow{5}{*}{\simcse} &  Rowenta RH8559 Delta Force 18V Cordless Bagless Energy Star Rated Stick Vacuum Cleaner \ldots \\\cline{2-2}
& Rowenta ZD100 Non-Toxic Soleplate Cleaner Kit \\\cline{2-2}
& Rowenta(R) Stainless Steel Soleplate Cleaning Kit ZD-110 \\\cline{2-2}
& Rowenta DR6015 Ultrasteam Hand-Held Steam Brush with Travel Pouch, 800-watt \\\cline{2-2}
& Rowenta DR6050 Ultrasteam Hand-Held Steam Brush Dual-Voltage with Travel Pouch, 800-watt \\\cline{1-2}

\multirow{5}{*}{\newreg} &  \textbf{Wrench} Set, Open End Metric 4mm-6mm - SCR-913.00 \\\cline{2-2}
& Craftsman 6 pc. Universal \textbf{Wrench} Set - Metric \\\cline{2-2}
& Tusk Spoke \textbf{Wrench} Set \\\cline{2-2}
& Crescent RD12BK 3/8-Inch Ratcheting Socket \textbf{Wrench} \\\cline{2-2}
& Allen \textbf{Wrench} Set, 10 Pc. Heavy Duty, Extra Long 9 T-handle, Metric Sizes \\\cline{1-2}

\hline
\end{tabular}
\caption{Top 5 predicted items for the query \textit{Rowe USA Spoke Wrench - Bagged 09-0001} given by various methods sorted by relevance. Correct items should be about \textit{wrench} and \newreg and \base model both give the same. Other models rely on spurious feature i.e. \textit{Rowe} for predicting items, which leads to wrong results}
\label{tab:qualitative_rowe}
\end{table*}

\begin{table*}
\small
\centering
\begin{tabular}{l|l}
\hline Method & Top 5 Predicted Items \\
\hline
\multirow{5}{*}{\base} & Camco 57723 Dust Cover for Portable Wave 6 Olympian \textbf{Heater} \\\cline{2-2}
& Sylvania SA200 10 Amp Outdoor Timer with Light Sensor \\\cline{2-2}
& Scosche CRAB Chrysler/Jeep Antenna Adapter \\\cline{2-2}
& Lasko 6435 Designer Series Ceramic Oscillating \textbf{Heater} with Remote Control \\\cline{2-2}
& Reusable Angel Ice Sculpture Mold \\\cline{1-2}

\multirow{5}{*}{\stdfine} & 3M Scotchlite Reflective Tape, Silver, 2-Inch by 36-Inch \\\cline{2-2}
& 3M Scotchlite Reflective Tape, Red, 2-Inch by 36-Inch \\\cline{2-2}
& Reflective Band - Made With Genuine Reflexite in America - By Jogalite (Pair of Two) \\\cline{2-2}
& Sunlite 4 Piece Bicycle Reflector Set with Brackets \\\cline{2-2}
& Road ID - Reflective Shoe Laces \\\cline{1-2}

\multirow{5}{*}{\oldreg} & 3M Scotchlite Reflective Tape, Silver, 1-Inch by 36-Inch \\\cline{2-2}
& 3M Scotchlite Reflective Tape, Red, 2-Inch by 36-Inch \\\cline{2-2}
& Reflective Band - Made With Genuine Reflexite in America - By Jogalite (Pair of Two) \\\cline{2-2}
& Casual Canine Reflective Jacket \\\cline{2-2}
& Aspects 264 Weather Dome \\\cline{1-2}

\multirow{5}{*}{\maskreg} & Reflective Band - Made With Genuine Reflexite in America - By Jogalite (Pair of Two) \\\cline{2-2}
& 3M Scotchlite Reflective Tape, Red, 2-Inch by 36-Inch \\\cline{2-2}
& Lasko 6435 Designer Series Ceramic Oscillating Heater with Remote Control \\\cline{2-2}
& 3M Scotchlite Reflective Tape, Silver, 1-Inch by 36-Inch \\\cline{2-2}
& Skylink PS-101 AAA+ Motion Sensor \\\cline{1-2}

\multirow{5}{*}{\simcse} & 3M Scotchlite Reflective Tape, Silver, 2-Inch by 36-Inch \\\cline{2-2}
& 3M Scotchlite Reflective Tape, Red, 2-Inch by 36-Inch \\\cline{2-2}
& Reflective Band - Made With Genuine Reflexite in America - By Jogalite (Pair of Two) \\\cline{2-2}
& Gates T274 Timing Belt \\\cline{2-2}
& Bell Automotive 22-5-00106-8 Heavy Duty Tubeless Tire Repair Kit \\\cline{1-2}

\multirow{5}{*}{\newreg} & Reflective Band - Made With Genuine Reflexite in America - By Jogalite (Pair of Two) \\\cline{2-2}
& Broan 679 Ventilation Fan and Light Combination \\\cline{2-2}
& 3M Scotchlite Reflective Tape, Silver, 1-Inch by 36-Inch \\\cline{2-2}
& 3M Scotchlite Reflective Tape, Red, 2-Inch by 36-Inch \\\cline{2-2}
& Roadpro 12V \textbf{Heater} and Fan with Swing-out Handle \\\cline{1-2}
\hline

\end{tabular}
\caption{Top 5 predicted items for the query \textit{Soleus Air Oscillating Reflective Heater} given by various methods sorted by relevance. Correct items should be about \textit{Heater} and \newreg and \base model both give the same. Other models rely on spurious feature i.e. \textit{Reflective} for predicting items, which leads to wrong results}
\label{tab:qualitative_reflective}
\end{table*}





\section{Complete Dataset Results}
\label{sec:complete_results}

For the complete dataset results, we follow the same setup as described before, with the only difference being number of epochs. We run \textit{Quora} for 4 epochs and \textit{SuperUser, Askubuntu} for 1 epoch. Results can be seen in Table~\ref{exp:quora_complete},\ref{exp:superuser_complete},\ref{exp:ubuntu_complete}.

\begin{table*}
\centering
\begin{tabular}{lllll}
\hline
\textbf{Pre-trained Model} & \textbf{Fine-tuning Method} & \textbf{Quora} & \textbf{AskUbuntu} & \textbf{Superuser}\\
\hline

\multirow{7}{*}{NLI-DistilBERT-Base} & \base & 14.00 & 31.80 & 48.50 \\\cline{2-5}  
& \stdfine & 54.84 $\pm$ 0.82 & 17.50 $\pm$ 2.41 & 29.25 $\pm$ 1.10 \\\cline{2-5} 
& \maskreg & 56.00 $\pm$ 1.15 & 19.79 $\pm$ 0.97 & 29.53 $\pm$ 1.59  \\\cline{2-5}
& \simcse & 56.24 $\pm$ 0.55 & 21.05 $\pm$ 1.57 & 29.04 $\pm$ 2.01  \\\cline{2-5}
& \oldreg & 52.69 $\pm$ 0.41 & 17.41 $\pm$ 1.16 & 27.33 $\pm$ 0.90  \\\cline{2-5}
& \newreg (0.15) & 57.19 $\pm$ 0.57 & 19.70 $\pm$ 1.33 & 28.67 $\pm$ 0.26  \\\cline{2-5}    
& \newreg (0.50) & 56.64 $\pm$ 0.74 & 20.56 $\pm$ 1.07 & 28.57 $\pm$ 1.04  \\\cline{1-5}    

\multirow{7}{*}{MSMARCO-DistilBERT-v4} & \base & 18.39 & 54.01  \\\cline{2-5}  
& \stdfine & 54.50 $\pm$ 0.31 & 19.79 $\pm$ 3.55 & 30.48 $\pm$ 1.33 \\\cline{2-5} 
& \maskreg & 56.00 $\pm$ 0.60 & 19.55 $\pm$ 3.12 & 33.71 $\pm$ 0.62 \\\cline{2-5}
& \simcse & 56.29 $\pm$ 1.62 & 19.57 $\pm$ 1.54 & 31.71 $\pm$ 0.93\\\cline{2-5}
& \oldreg & 54.41 $\pm$ 0.04 & 32.01 $\pm$ 1.72 & 46.80 $\pm$ 1.75 \\\cline{2-5}
& \newreg (0.15) & 56.63 $\pm$ 0.65 & 20.09 $\pm$ 1.46 & 33.06 $\pm$ 1.65 \\\cline{2-5}    
& \newreg (0.50) & 55.98 $\pm$ 0.68 & 19.50 $\pm$ 1.77 & 33.51 $\pm$ 0.62 \\\cline{1-5}    



\hline
\end{tabular}
\caption{\label{exp:quora_complete}
Trained on Complete \textit{Quora} for 4 epochs (same hyperparameters as in paper)}
\end{table*}

\begin{table*}
\centering
\begin{tabular}{lllll}
\hline
\textbf{Pre-trained Model} & \textbf{Fine-tuning Method} & \textbf{Quora} & \textbf{AskUbuntu} & \textbf{Superuser}\\
\hline

\multirow{7}{*}{NLI-DistilBERT-Base} & \base & 14.00 & 31.80 & 48.50 \\\cline{2-5}  
& \stdfine & 11.02 $\pm$ 0.43 & 71.09 $\pm$ 1.05 & 74.90 $\pm$ 0.27\\\cline{2-5} 
& \maskreg & 13.36 $\pm$ 0.38 & 77.62 $\pm$ 0.31 & 79.09 $\pm$ 0.21  \\\cline{2-5}
& \simcse & 12.41 $\pm$ 0.34 & 74.73 $\pm$ 0.79 & 77.58 $\pm$ 0.11 \\\cline{2-5}
& \oldreg & 14.75 $\pm$ 0.41 & 68.18 $\pm$ 0.64 & 72.88 $\pm$ 1.14 \\\cline{2-5}
& \newreg (0.15) & 13.90 $\pm$ 0.51 & 77.54 $\pm$ 0.33 & 79.70 $\pm$ 0.31  \\\cline{2-5}    
& \newreg (0.50) & 13.77 $\pm$ 0.63 & 77.75 $\pm$ 0.49 & 81.06 $\pm$ 0.32  \\\cline{1-5}    


\multirow{7}{*}{MSMARCO-DistilBERT-v4} & \base & 18.39 & 54.01 & 80.73 \\\cline{2-5}  
& \stdfine & 12.01 $\pm$ 0.33 & 70.70 $\pm$ 1.16 & 74.94 $\pm$ 0.22 \\\cline{2-5} 
& \maskreg & 13.66 $\pm$ 0.28 & 78.04 $\pm$ 0.44 & 79.94 $\pm$ 0.26  \\\cline{2-5}
& \simcse & 13.89 $\pm$ 0.37 & 75.39 $\pm$ 0.99 & 78.53 $\pm$ 0.15 \\\cline{2-5}
& \oldreg & 17.34 $\pm$ 0.26 & 78.84 $\pm$ 0.49 & 85.01 $\pm$ 1.41  \\\cline{2-5}
& \newreg (0.15) & 14.84 $\pm$ 0.53 & 78.10 $\pm$ 0.55 & 81.19 $\pm$ 0.67 \\\cline{2-5}    
& \newreg (0.50) & 15.20 $\pm$ 0.41 & 78.47 $\pm$ 0.17 & 82.27 $\pm$ 0.36 \\\cline{1-5}    

\hline
\end{tabular}
\caption{\label{exp:ubuntu_complete}
Trained on Complete \textit{AskUbuntu} for 1 epoch (same hyperparameters as in paper)}
\end{table*}

\begin{table*}
\centering
\begin{tabular}{lllll}
\hline
\textbf{Pre-trained Model} & \textbf{Fine-tuning Method} & \textbf{Quora} & \textbf{AskUbuntu} & \textbf{Superuser} \\
\hline

\multirow{7}{*}{NLI-DistilBERT-Base} & \base & 14.00 & 31.80 & 48.50 \\\cline{2-5}  
& \stdfine & 12.68 $\pm$ 0.37 & 50.90 $\pm$ 2.95 & 83.50 $\pm$ 0.97\\\cline{2-5} 
& \maskreg & 13.65 $\pm$ 0.31 & 40.56 $\pm$ 2.66 & 87.79 $\pm$ 0.67 \\\cline{2-5}
& \simcse & 14.39 $\pm$ 0.64 & 53.64 $\pm$ 1.67 & 85.85 $\pm$ 0.58  \\\cline{2-5}
& \oldreg & 15.60 $\pm$ 0.13 & 35.41 $\pm$ 2.91 & 80.68 $\pm$ 0.73 \\\cline{2-5}
& \newreg (0.15) & 15.19 $\pm$ 0.20 & 44.38 $\pm$ 1.10 & 88.31 $\pm$ 0.65 \\\cline{2-5}    
& \newreg (0.50) & 14.30 $\pm$ 0.18 & 45.83 $\pm$ 5.42 & 88.78 $\pm$ 0.39  \\\cline{1-5}    


\multirow{7}{*}{MSMARCO-DistilBERT-v4} & \base & 18.39 & 54.01 & 80.73 \\\cline{2-5}  
& \stdfine & 12.61 $\pm$ 0.46 & 52.32 $\pm$ 3.07 & 84.60 $\pm$ 0.77 \\\cline{2-5} 
& \maskreg & 14.20 $\pm$ 0.22 & 38.32 $\pm$ 4.16 & 88.63 $\pm$ 0.44 \\\cline{2-5}
& \simcse & 14.08 $\pm$ 0.39 & 49.04 $\pm$ 3.56 & 86.55 $\pm$ 0.32 \\\cline{2-5}
& \oldreg & 17.42 $\pm$ 0.40 & 57.83 $\pm$ 1.31 & 89.94 $\pm$ 0.23 \\\cline{2-5}
& \newreg (0.15) & 16.03 $\pm$ 0.66 & 52.77 $\pm$ 2.07 & 88.85 $\pm$ 0.43 \\\cline{2-5}    
& \newreg (0.50) & 16.24 $\pm$ 0.31 & 56.33 $\pm$ 1.91 & 89.90 $\pm$ 0.30  \\\cline{1-5}    

\hline
\end{tabular}
\caption{\label{exp:superuser_complete}
Trained on Complete \textit{SuperUser} for 1 epoch (same hyperparameters as in paper)}
\end{table*}

\section{10\% Dataset Subset Results}
\label{sec:subset_results}

Results on 10\% subset of the data (most of them are redundant as they overlap with results in main paper) can be seen here: Table~\ref{exp:quora_10},\ref{exp:superuser_10},\ref{exp:ubuntu_10}.

\begin{table*}
\centering
\begin{tabular}{lllll}
\hline
\textbf{Pre-trained Model} & \textbf{Fine-tuning Method} & \textbf{Quora} & \textbf{AskUbuntu} & \textbf{Superuser}\\
\hline

\multirow{7}{*}{NLI-DistilBERT-Base} & \base & 14.00 & 31.80 & 48.50 \\\cline{2-5}  
& \stdfine & 33.22 $\pm$ 0.82 & 10.95 $\pm$ 1.45 & 18.56 $\pm$ 1.53 \\\cline{2-5} 
& \maskreg & 35.02 $\pm$ 0.85 & 12.94 $\pm$ 1.05 & 21.48 $\pm$ 1.65 \\\cline{2-5}
& \simcse & 32.52 $\pm$ 0.29 & 10.82 $\pm$ 1.19 & 20.99 $\pm$ 1.13 \\\cline{2-5}
& \oldreg & 33.41 $\pm$ 0.70 & 14.91 $\pm$ 1.26 & 25.63 $\pm$ 1.67\\\cline{2-5}
& \newreg (0.15) & 33.94 $\pm$ 1.47 & 13.73 $\pm$ 0.66 & 22.74 $\pm$ 0.59 \\\cline{2-5}    
& \newreg (0.50) & 32.93 $\pm$ 0.50 & 12.33 $\pm$ 0.89 & 20.91 $\pm$ 0.52 \\\cline{1-5}    

\multirow{7}{*}{MSMARCO-DistilBERT-v4} & \base & 18.39 & 54.01 & 80.73  \\\cline{2-5}  
& \stdfine & 34.15 $\pm$ 0.30 & 13.31 $\pm$ 1.29 & 23.73 $\pm$ 1.63 \\\cline{2-5} 
& \maskreg & 36.49 $\pm$ 0.70 & 11.89 $\pm$ 1.70 & 25.88 $\pm$ 2.49  \\\cline{2-5}
& \simcse & 34.85 $\pm$ 1.32 & 11.34 $\pm$ 1.99 & 25.63 $\pm$ 0.64 \\\cline{2-5}
& \oldreg & 38.50 $\pm$ 0.29 & 31.96 $\pm$ 3.23 & 55.98 $\pm$ 1.49 \\\cline{2-5}
& \newreg (0.15) & 38.41 $\pm$ 1.09 & 15.91 $\pm$ 2.22 & 30.95 $\pm$ 1.49 \\\cline{2-5}    
& \newreg (0.50) & 35.26 $\pm$ 1.09 & 12.28 $\pm$ 1.58 & 24.61 $\pm$ 0.95  \\\cline{1-5}    

\hline
\end{tabular}
\caption{\label{exp:quora_10}
Trained on 10\% \textit{Quora}}
\end{table*}

\begin{table*}
\centering
\begin{tabular}{lllll}
\hline
\textbf{Pre-trained Model} & \textbf{Fine-tuning Method} & \textbf{Quora} & \textbf{AskUbuntu} & \textbf{Superuser}\\
\hline

\multirow{7}{*}{NLI-DistilBERT-Base} & \base & 14.00 & 31.80 & 48.50 \\\cline{2-5}  
& \stdfine & 12.74 $\pm$ 0.43 & 65.68 $\pm$ 0.54 & 72.70 $\pm$ 0.73 \\\cline{2-5} 
& \maskreg & 12.44 $\pm$ 0.48 & 71.51 $\pm$ 1.16 & 77.55 $\pm$ 0.57 \\\cline{2-5}
& \simcse & 12.87 $\pm$ 0.36 & 67.31 $\pm$ 1.28 & 73.46 $\pm$ 2.24 \\\cline{2-5}
& \oldreg & 14.09 $\pm$ 0.18 & 56.86 $\pm$ 0.69 & 66.28 $\pm$ 0.66 \\\cline{2-5}
& \newreg (0.15) & 13.62 $\pm$ 0.25 & 71.45 $\pm$ 1.17 & 78.32 $\pm$ 0.58 \\\cline{2-5}    
& \newreg (0.50) & 13.57 $\pm$ 0.32 & 71.24 $\pm$ 0.62 & 78.89 $\pm$ 0.59  \\\cline{1-5}    


\multirow{7}{*}{MSMARCO-DistilBERT-v4} & \base & 18.39 & 54.01 & 80.73 \\\cline{2-5}  
& \stdfine & 14.05 $\pm$ 0.19 & 68.11 $\pm$ 1.09 & 75.33 $\pm$ 0.74\\\cline{2-5} 
& \maskreg & 12.63 $\pm$ 0.36 & 72.59 $\pm$ 0.41 & 79.00 $\pm$ 0.08 \\\cline{2-5}
& \simcse & 14.98 $\pm$ 0.08 & 70.07 $\pm$ 0.32 & 76.54 $\pm$ 1.43\\\cline{2-5}
& \oldreg & 17.16 $\pm$ 0.13 & 73.04 $\pm$ 0.77 & 84.09 $\pm$ 0.28 \\\cline{2-5}
& \newreg (0.15) & 15.67 $\pm$ 0.22 & 74.11 $\pm$ 0.58 & 81.78 $\pm$ 0.28 \\\cline{2-5}    
& \newreg  (0.50) & 16.11 $\pm$ 0.36 & 74.64 $\pm$ 0.56 & 82.86 $\pm$ 0.65 \\\cline{1-5}    

\hline
\end{tabular}
\caption{\label{exp:ubuntu_10}
Trained on 10 \% \textit{AskUbuntu}}
\end{table*}

\begin{table*}
\centering
\begin{tabular}{lllll}
\hline
\textbf{Pre-trained Model} & \textbf{Fine-tuning Method} & \textbf{Quora} & \textbf{AskUbuntu} & \textbf{Superuser}\\
\hline

\multirow{7}{*}{NLI-DistilBERT-Base} & \base & 14.00 & 31.80 & 48.50  \\\cline{2-5}  
& \stdfine & 12.82 $\pm$ 0.49 & 47.98 $\pm$ 6.63 & 75.70 $\pm$ 1.49 \\\cline{2-5} 
& \maskreg & 13.13 $\pm$ 0.17 & 33.83 $\pm$ 3.07 & 81.97 $\pm$ 0.33\\\cline{2-5}
& \simcse & 13.41 $\pm$ 0.20 & 45.03 $\pm$ 4.22 & 76.80 $\pm$ 0.68  \\\cline{2-5}
& \oldreg & 14.44 $\pm$ 0.28 & 36.52 $\pm$ 1.53 & 71.11 $\pm$ 0.86 \\\cline{2-5}
& \newreg (0.15) & 14.35 $\pm$ 0.28 & 40.10 $\pm$ 3.68 & 82.27 $\pm$ 0.38 \\\cline{2-5}    
& \newreg (0.50) &  14.67 $\pm$ 0.18 & 47.00 $\pm$ 1.58 & 83.40 $\pm$ 0.54 \\\cline{1-5}    


\multirow{7}{*}{MSMARCO-DistilBERT-v4} & \base & 18.39 & 54.01\\\cline{2-5}  
& \stdfine & 14.35 $\pm$ 0.20 & 59.27 $\pm$ 2.16 & 79.28 $\pm$ 0.94\\\cline{2-5} 
& \maskreg & 13.65 $\pm$ 0.20 & 32.05 $\pm$ 6.53 & 83.53 $\pm$ 0.53 \\\cline{2-5}
& \simcse & 14.70 $\pm$ 0.05 & 48.97 $\pm$ 3.28 & 80.51 $\pm$ 0.70 \\\cline{2-5}
& \oldreg &  17.64 $\pm$ 0.25 & 60.75 $\pm$ 0.84 & 86.24 $\pm$ 0.13 \\\cline{2-5}
& \newreg (0.15) & 15.94 $\pm$ 0.15 & 53.02 $\pm$ 5.42 & 85.44 $\pm$ 0.23 \\\cline{2-5}    
& \newreg (0.50) & 16.39 $\pm$ 0.31 & 60.07 $\pm$ 2.42 & 86.24 $\pm$ 0.24 \\\cline{1-5}    

\hline
\end{tabular}
\caption{\label{exp:superuser_10}
Trained on 10\% \textit{SuperUser}}
\end{table*}

\section{Experimental Budget}

\paragraph{Computing Infrastructure} We use 16GB V100 GPUs for all our experiments

\paragraph{Training Time} Our training time is around 1 hour for each run on Question Recommendation. Amazon dataset takes ~8 hrs for training.

\paragraph{Parameters Of Model} We use DistilBERT model for all our experiments.

\begin{table*}
\footnotesize
\centering
\resizebox{\textwidth}{!}{%
\begin{tabular}{ll|lll|lll|lll}
\hline
\textbf{\multirow{2}{*}{Base Model}} & \textbf{\multirow{2}{*}{Method}} & \multicolumn{3}{c|}{Train : Quora} & \multicolumn{3}{c}{Train : AskUbuntu} & \multicolumn{3}{c}{Train : Superuser} \\
& & \textbf{Quora} & \textbf{AskUbuntu} & \textbf{Superuser}  & \textbf{Quora} & \textbf{AskUbuntu} & \textbf{Superuser}& \textbf{Quora} & \textbf{AskUbuntu} & \textbf{Superuser} \\
\hline

\multirow{6}{*}{\small{NLI}} & \base & 0.140 $\pm$ 0.000 & 0.318 $\pm$ 0.000 & 0.486 $\pm$ 0.000 & 0.140 $\pm$ 0.000 & 0.318 $\pm$ 0.000 & 0.486 $\pm$ 0.000 & 0.140 $\pm$ 0.000 & 0.318 $\pm$ 0.000 & 0.486 $\pm$ 0.000 \\\cline{2-11}

& \stdfine & 0.559 $\pm$ 0.016 & 0.181 $\pm$ 0.018 & 0.257 $\pm$ 0.021 & 0.092 $\pm$ 0.001 & 0.775 $\pm$ 0.000 & 0.762 $\pm$ 0.002 & 0.122 $\pm$ 0.014 & 0.632 $\pm$ 0.029 & 0.863 $\pm$ 0.005 \\\cline{2-11}

& \maskreg & 0.529 $\pm$ 0.046 & 0.231 $\pm$ 0.014 & 0.331 $\pm$ 0.031 & 0.119 $\pm$ 0.010 & 0.810 $\pm$ 0.008 & 0.803 $\pm$ 0.010 & 0.132 $\pm$ 0.000 & 0.417 $\pm$ 0.057 & 0.901 $\pm$ 0.007 \\\cline{2-11}

& \simcse & 0.571 $\pm$ 0.003 & 0.176 $\pm$ 0.023 & 0.259 $\pm$ 0.004 & 0.118 $\pm$ 0.012 & 0.792 $\pm$ 0.001 & 0.794 $\pm$ 0.012 & 0.137 $\pm$ 0.003 & 0.587 $\pm$ 0.024 & 0.876 $\pm$ 0.005 \\\cline{2-11}

& \oldreg & 0.553 $\pm$ 0.002 & 0.149 $\pm$ 0.015 & 0.225 $\pm$ 0.005 & 0.146 $\pm$ 0.003 & 0.691 $\pm$ 0.009 & 0.737 $\pm$ 0.003 & 0.155 $\pm$ 0.000 & 0.383 $\pm$ 0.014 & 0.777 $\pm$ 0.019 \\\cline{2-11}

& \newreg & 0.566 $\pm$ 0.013 & 0.186 $\pm$ 0.011 & 0.295 $\pm$ 0.018 & 0.136 $\pm$ 0.001 & 0.820 $\pm$ 0.001 & 0.815 $\pm$ 0.001 & 0.151 $\pm$ 0.000 & 0.563 $\pm$ 0.015 & 0.903 $\pm$ 0.005 \\\cline{2-11}

\hline

\multirow{6}{*}{\small{MSMARCO}} & \base & 0.184 $\pm$ 0.000 & 0.540 $\pm$ 0.000 & 0.807 $\pm$ 0.000 & 0.184 $\pm$ 0.000 & 0.540 $\pm$ 0.000 & 0.807 $\pm$ 0.000 & 0.184 $\pm$ 0.000 & 0.540 $\pm$ 0.000 & 0.807 $\pm$ 0.000 \\\cline{2-11}

& \stdfine & 0.564 $\pm$ 0.001 & 0.176 $\pm$ 0.005 & 0.265 $\pm$ 0.001 & 0.104 $\pm$ 0.007 & 0.776 $\pm$ 0.013 & 0.757 $\pm$ 0.011 & 0.144 $\pm$ 0.011 & 0.601 $\pm$ 0.020 & 0.862 $\pm$ 0.009 \\\cline{2-11}

& \maskreg & 0.526 $\pm$ 0.008 & 0.225 $\pm$ 0.011 & 0.363 $\pm$ 0.017 & 0.108 $\pm$ 0.004 & 0.813 $\pm$ 0.004 & 0.810 $\pm$ 0.000 & 0.131 $\pm$ 0.005 & 0.434 $\pm$ 0.007 & 0.908 $\pm$ 0.005 \\\cline{2-11}

& \simcse & 0.574 $\pm$ 0.019 & 0.187 $\pm$ 0.006 & 0.301 $\pm$ 0.017 & 0.123 $\pm$ 0.004 & 0.797 $\pm$ 0.005 & 0.787 $\pm$ 0.007 & 0.125 $\pm$ 0.004 & 0.575 $\pm$ 0.050 & 0.887 $\pm$ 0.000 \\\cline{2-11}

& \oldreg & 0.558 $\pm$ 0.005 & 0.297 $\pm$ 0.022 & 0.438 $\pm$ 0.006 & 0.166 $\pm$ 0.001 & 0.805 $\pm$ 0.002 & 0.851 $\pm$ 0.001 & 0.173 $\pm$ 0.004 & 0.614 $\pm$ 0.013 & 0.903 $\pm$ 0.001 \\\cline{2-11}

& \newreg & 0.577 $\pm$ 0.002 & 0.175 $\pm$ 0.001 & 0.286 $\pm$ 0.011 & 0.134 $\pm$ 0.004 & 0.831 $\pm$ 0.004 & 0.826 $\pm$ 0.000 & 0.159 $\pm$ 0.002 & 0.575 $\pm$ 0.000 & 0.911 $\pm$ 0.008 \\\cline{2-11}

\hline







\end{tabular}
}
\caption{\label{exp:rerun_main_table}
AUC (0.05) using two different base models, NLI-DistilBERT-Base and MSMARCO-DistilBERT-Base. First three columns correspond to training on \textit{Quora}, middle three to \textit{AskUbuntu} and the  last three  training on \textit{SuperUser}. When evaluated on technical forums with a weaker base model (NLI), \newreg obtains the best AUC on both IID and OOD evaluation. These are different from Table \ref{exp:main_table}, they run with lower learning rate of 1e-5 and higher number of epochs with best model numbers reported }
\end{table*}


\end{document}